\documentclass[3p]{elsarticle}

\usepackage{geometry}
\usepackage{graphicx}
\usepackage{amsmath} 
\usepackage{amssymb}  
\usepackage{amsfonts}
\usepackage{subfigure}
\usepackage[utf8]{inputenc}
\usepackage{graphicx, multicol, latexsym, amsmath, amssymb}
\usepackage[utf8]{inputenc}
\usepackage[T1]{fontenc}

\usepackage{epsfig} 
\usepackage{amsmath} 
\usepackage{amssymb}  
\usepackage{amsfonts}
\usepackage[export]{adjustbox}
\usepackage[subnum]{cases}
\usepackage{graphicx, multicol, latexsym, amsmath, amssymb}
\usepackage{multirow}
\usepackage{textcomp}
\usepackage[subnum]{cases}
\usepackage{matlab-prettifier}
\usepackage{listings}
\usepackage{eucal}
\usepackage{algorithmic}
\usepackage[linesnumbered,titlenumbered,lined,titlenumbered]{algorithm2e}
\usepackage{float}
\usepackage{caption}
\usepackage{comment}
\usepackage{braket}
\usepackage{booktabs}
\usepackage{siunitx}
\usepackage{verbatim}
\usepackage{afterpage}
\usepackage{mathrsfs}

\usepackage{xcolor}
\usepackage{comment}

\makeatletter
\gdef\emailauthor#1#2{\stepcounter{ead}%
     \g@addto@macro\@elseads{\raggedright%
      \let\corref\@gobble
      \eadsep\texttt{#1}\def\eadsep{\unskip,\space}}%
}
\def\urlauthor#1#2{\g@addto@macro\@elsuads{\let\corref\@gobble%
    \raggedright\eadsep\texttt{#1}%
    \def\eadsep{\unskip,\space}}%
}
\makeatother

\biboptions{sort&compress}

\journal{Nuclear Fusion}

\usepackage{hyperref}
\hypersetup{
  breaklinks=true,
  pdfborder={0 0 0}
}

\begin{document}

\begin{frontmatter}


\title{Applications of a novel model-based real-time observer for electron density profile control experiments in TCV}



\author[epfl]{F.~Pastore}\ead{francesco.pastore@epfl.ch}    
\author[epfl]{O.~Sauter}
\author[deepmind]{F.~Felici}
\author[ctu,ipp_germany]{D.~Kropackova}
\author[epfl]{A.~Balestri}
\author[epfl]{C.~Galperti}
\author[ipp_germany]{O.~Kudlacek}
\author[epfl]{K.~Lee}
\author[epfl]{A.~Pau}
\author[iter]{T.~Ravensbergen}
\author[iter]{S.~Van Mulders}
\author[epfl]{B.~Vincent}
\author[iter]{N.M.T.~Vu}

\author[tcv]{the TCV team}
\author[wpte]{ and the EUROfusion Tokamak Exploitation Team.}

\address[epfl]{École Polytechnique Fédérale de Lausanne (EPFL), Swiss Plasma Center (SPC), CH-1015 Lausanne, Switzerland}
\address[deepmind]{Google DeepMind, London, UK. }
\address[ctu]{Czech Technical University, Faculty of Nuclear Sciences and Physical Engineering, Brehova 7, 115 19, Prague 1, Czech Republic.}
\address[ipp_germany]{Max-Planck-Institut für Plasmaphysik, 85748 Garching bei München, Germany.}
\address[iter]{ITER Organization, Route de Vinon-sur-Verdon, 13067 St. Paul Lez Durance Cedex, France.}
\address[tcv]{See the author list of B. Duval et al., 2024 Nucl. Fusion 64 112023}
\address[wpte]{See the author list of E. Joffrin et al., 2024 Nucl. Fusion in press https://doi.org/10.1088/1741-4326/ad2be4 }

\begin{abstract}

 The main aspects of real-time control of tokamak plasmas today are ensuring a stationary state around a highly performing target, proximity control to avoid disruptions, and controlling the access (ramp-up) and termination (ramp-down) of the shot. Within these applications, the real-time estimation and control of the electron density is fundamental for monitoring plasma confinement, heating efficiency, and control exhaust conditions, impurity concentration, fusion power production, and proximity to the so-called density limit \cite{Greenwald_1988}. Building on the integration of a novel multi-rate observer \cite{PASTORE2023113615} based on RAPDENS \cite{Blanken2018} into the TCV plasma control system, this follow-up study explores its application to density profile control for detachment studies, ECH- and NBH- L-mode plasmas, and high-performance H-mode scenarios.


Experiments on TCV demonstrate the observer’s capability to support detachment studies in complex divertor geometries by accurately controlling the line-averaged density within the last-closed flux surface while avoiding the pick-up of the far-infrared interferometer signal attributed to the Scrape-Off Layer electron density in the divertor region.
The estimated electron density profile from the observer enables local control of the central density in an L-mode plasma with a mix of ECH and NBH, to operate below cutoff conditions. The effect of these heating schemes on the electron density profile peaking is analyzed and treated as a disturbance to the control task. Real-time estimation and adoption of the unknown, time-varying particle transport coefficient profiles, such as the ratio between the electron pinch velocity and diffusion coefficient, is carried out to further improve the predictive capabilities and spatial accuracy of the control-oriented model. The turbulent transport observed in the experimental scenario is characterized using linear and non-linear gyrokinetic simulations with the GENE code to verify the observed particle pumpout in these experiments. Relying on the good spatial resolution of the density profile reconstruction at the top of the pedestal, the simultaneous control of the edge-normalized density and toroidal beta of high-performance H-mode plasmas is demonstrated. This approach yields good confinement properties, scenario reproducibility, and an edge-density metric independent of the diagnostics available on the tokamak, while avoiding density limits, plasma disruptions, and the propagation of diagnostic faults in the control scheme.

\end{abstract}


\end{frontmatter}


\section{Introduction \& Motivation}\label{section:Introduction}
Estimation and control of plasma kinetic quantities, such as plasma electron density, temperature, and current density profile, have gained increased relevance over the years. To maximize plasma fusion output, avoid operational limits and MHD instabilities, and keep the plasma scenario controlled over stable trajectories, a precise real-time assessment of these quantities is deemed necessary \cite{Snipes2017,BIEL20158}. Future fusion power plants represent a challenging environment for many physics, technological, and engineering problems. The nominal operation point of the power plant is bound to lie close to physical and technological limits for an economical scalability \cite{Ward2005_EconomicViability} and electric cost competitiveness. The operation of the fusion reactor is also constrained by a limited set of diagnostics, due to the harsh reactor conditions expected on a burning plasma device. The Plasma Control System (PCS) of the reactor has to be designed to guarantee robust and reliable operation, integrate and orchestrate the different signals coming from diagnostics, actuators, and algorithms, handle off-normal events, and ensure robust performance in real time in case of diagnostics failures or unavailability. 

The integration of plasma state observers in the PCS, designed to predict and reconstruct the plasma state, can be exploited to provide robust performance by combining different diagnostics and models in real time in an integrated fashion. The reconstructed state can then inform the supervisor and off-normal event manager (e.g. SAMONE \cite{Vu_2021}) and track different plasma quantities to ensure that the operational limits are not exceeded and lie in a high-performance operational space.

\subsection{State of the art - kinetic profiles estimators for offline and real-time applications}

There are multiple examples of observers developed in recent years for offline and real-time plasma kinetic state estimation. These observers perform the reconstruction of plasma kinetic profiles, exploiting the available information provided by diagnostics, equilibrium codes, and actuator signals.

Model-based approaches, relying on the solution of flux-surface averaged equations for the evolution of kinetic profiles, have been developed and tested on different plasma devices over the last decade \cite{Felici2014_ACC, Morishita2022_ASTI,Citrin2024_TORAX,Pajares2022_IntegratedScalars_NTMSuppression, Morosohk2025, PASTORE2023113615, Kropackova2025,VanMulders2025,VanMulders2025_TCV}. Data-driven approaches, based on machine learning techniques, have been recently designed and deployed on fusion machines for the real-time reconstruction of kinetic quantities \cite{Shousha2024_RTCAKENN,Jalalvand2022_ReservoirRC,kim_real-time_2025} and the enhancement of diagnostic data \cite{Illerhaus2025_GPUpipeline}.

A model-based approach to predict and estimate the electron density profile has been introduced in \cite{Blanken2018}. RAPDENS, the RApid Plasma DENsity Simulator, has been formulated, programmed, and tested in real time on TCV and ASDEX Upgrade. It incorporates a control-oriented model to describe the dynamics of the flux-surface averaged electron density profile in a tokamak. The modeling of the neutrals in the vacuum chamber and reactor wall has been introduced with two dedicated 0-D inventories to describe the ionization, recycling, recombination, flux of particles to the Scrape-Off Layer (SOL) and plasma fueling through different actuators, such as gas puffing, Neutral Beam Injection (NBI), and pellet fueling. The real-time-capable state observer is based on the Extended Kalman Filter (EKF) \cite{Kalman1960} technique, where diagnostic measurements related to the electron density or neutrals are incorporated into the prediction of the plasma state provided by the non-linear physics model. A first real-time application of the code on TCV and ASDEX Upgrade has been shown in \cite{Blanken2019}, where integral measurements of interfermeters and Bremsstrahlung are integrated in the EKF.

RAPDENS follows a similar approach as the RAPTOR code \cite{felici_real-time_2011}, the RApid Plasma Transport simulatOR, developed to predict and estimate the plasma density, temperature, and current density profiles.
Recent results on the estimation of the electron temperature $T_{e}$, density $n_e$, and current density profiles for ITER \cite{VanMulders2025} using a synthetic diagnostic model for Thomson scattering (TS) system have been proposed, where the simultaneous estimation of the plasma state and unknown plasma parameters has been carried out within the formulation of RAPTOR EKF. The improved state estimator is then leveraged for offline Kinetic Equilibrium Reconstruction of TCV plasmas \cite{VanMulders2025_TCV}, where a self-consistent solution of the magnetic equilibrium problem and internal kinetic state is achieved, leading to accurate reconstruction of the safety factor profile and dynamical evolution of the sawtooth instability. 


In \cite{PASTORE2023113615}, a novel multi-rate EKF based on the RAPDENS model has been presented, fusing high-frequency far-infrared interferometer (FIR) data and low-frequency TS data to provide an estimate of the electron density profile with high spatial and temporal accuracy.
The model-based observer, integrated in the TCV PCS \cite{GALPERTI2024114640}, has been further refined and exploited to assist various experimental missions that can benefit from direct control of the electron density profile. These experiments are used to validate the proposed framework for high-performance, high-density plasmas, with various heating sources, as well as demonstrating how real-time density profile control can benefit physics studies.

\subsection{Leveraging model-based observers for density estimation, control, and synthesis of model-based controllers}
The estimated plasma density profile provides valuable information to the density controller(s) and plasma state monitor to achieve the required tailored density profile for a given plasma scenario. 
Profile control of the electron density is useful, for example, in establishing the plasma density conditions for full absorption of injected microwave power for ECH and ECCD operation below the cutoff limit.
Real-time reconstruction of the density profile enables, in addition, the tracking and estimation of the microwave beam tracing \cite{poli_torbeam_2018} and Neutral Beam Injection (NBI) power deposition profile \cite{weiland_real-time_2023}.
Control of the plasma edge density in H-mode plasmas is required to avoid HL back-transitions or disruptions due to the density limit \cite{Greenwald_1988}, while operating at high densities to optimize the fusion gain. Profile observers enable the tracking of the plasma electron edge density in high-density, high-performance shots for proximity control or disruption avoidance experiments. Tailoring the density profile with a combination of different fueling sources, such as pellet injection and gas puffing \cite{Orrico2025}, can be beneficial to avoid tungsten impurity accumulation \cite{BosmanITER}, thus increasing the fusion efficiency by minimizing the radiated power in the core, and preventing fuel dilution. 

The density state observer provides not only a more comprehensive information in real time of the density profile state, but it can be leveraged to detect, correct and/or discard corrupted diagnostics channels, such as interferometers fringe jumps \cite{Blanken2018,Bosman2021,Murari2006}, and effectively decouple the diagnostics system from the controllers, thus improving the reliability of the overall control system and resulting in a machine-agnostics representation of the plasma state \cite{Vu_2021}. 

The fast capability of the code, condensing in a control-oriented manner the particle physics of a tokamak in its predictive model, can be exploited to design and test feedforward and feedback controllers for ramp-up, flat-top, and ramp-down applications, in view of the operation of future fusion reactors such as ITER \cite{Ravensbergen_2017}.

\subsection{Application of the multi-rate density observer on TCV}
In this article, the main results achieved by leveraging the novel multi-rate electron density observer are summarized, validating the procedure in various scenarios and demonstrating how the improved observer can aid in estimating the density state compared to the traditional density control system routinely employed for operation on TCV.

Real-time estimation of the density profile provided by the RAPDENS observer enables the control of derived quantities of the reconstructed profile, such as local control of the electron density profile on a specific profile point, or the line-averaged electron density computed within the Last-Closed Flux Surface (LCFS). The use of the density profile observer has been carried out with Single-Input Single-Output (SISO) density control experiments, by propagating the reconstructed reference signal of the $n_e$ profile to an anti-windup PI controller for the gas flux fueling modulation.

 The control of the density upstream conditions in detachment studies in support of alternative divertor studies has been enabled with the real-time estimation of the line-averaged electron density enclosed within the LCFS. The pick-up of density in the SOL is avoided with the combined information provided by the real-time mapping of the TS points in the poloidal plane with the available FIR measurements.
 
Local control of central electron density in ECH plasmas for operation below cutoff is demonstrated with different mixes of heating schemes.

The simultaneous control of the plasma edge density and $\mathrm{\beta_{\mathrm{tor}}}$ in H-mode plasmas has been carried out, leading to a stable, robust, and repeatable scenario to study disruption avoidance and proximity control schemes in high-density plasma conditions (normalized beta $\mathrm{\beta_N \approx 2.15}$, Greenwald fraction \cite{Greenwald_1988} $\mathrm{f_{\mathrm{GW}} \approx 0.80}$).

The estimation of the unknown plasma transport coefficient profile in real time,  the electron pinch velocity, is obtained by the observer and actively employed during the experiments, based on the information of the Real Time (RT)-TS data in the EKF procedure \cite{pastoreEPS}. The estimated pinch transport coefficient is then employed in the RAPDENS predictive model to increase the spatial accuracy of the reconstruction with the latest information provided by the TS diagnostic system. This represents a significant enhancement of the RAPDENS model for real-time predictive capabilities, in complement to imposing a more reliable boundary condition at the plasma edge \cite{Kropackova2025}.

\subsection{Structure of the work}
The structure of the remaining part of this work is as follows: in Section \ref{section:RAPDENS equations}, a summary of the underlying equations solved in the novel multi-rate density observer and the estimation of the unknown electron pinch coefficient is presented. Section \ref{section:Control scheme SCD} shows in detail the integration of the electron density observer in the TCV PCS, where the digital density profile controller, using the information contained in the reconstructed electron density profile, is deployed. Section \ref{section:Detachment} covers the experimental results on the upstream density control in support of detachment studies, where the magnetic equilibrium divertor geometry impacts the traditional density control feedback loop, due to the non-negligible pickup of the SOL density in the divertor region. 
Section \ref{section: ne control below cutoff} expands on the results related to SISO control of the local (in radius) electron density, using the accurate profile estimate, in ECH shots, for operation below cutoff. The experiments are conducted in the presence of NBI in co- and counter-current direction as a disturbance to the control task. The effects of the different actuators on the shape of the density profile are analyzed qualitatively with the RAPDENS profile reconstruction in various time intervals of the shot. A quantitative analysis of the observed particle pumpout effect \cite{Weisen2001} has been carried out with linear and non-linear GENE \cite{GENE} flux-tube simulations.

The adoption of the upgraded density feedback loop to control the edge density is presented in Section \ref{section: ne edge control H-mode}, where the simultaneous control of edge electron density and beta is carried out for high-density, high-performance H-mode shots. The multi-rate observer, which employs TS and only a subset of the available FIR signals in real time, has been capable of tracking the edge density with precision comparable to that of the offline TS data. The recovery and reconstruction of FIR signals corrupted by fringe jumps is also showcased, improving the robustness of the TCV control system and the reconstruction of the density profile in high-density scenarios that are prone to this problem.

The conclusions and outlook of this work are finally reported in Section \ref{section: conclusion}.
\section{Formulation of the multi-rate electron density profile observer}
\label{section:RAPDENS equations}

\subsection{Summary of the equations of the electron density observer}\label{section:summary_RAPDENS_EKF_equations}
The electron density observer has been designed, tuned, and tested with offline TCV data in \cite{PASTORE2023113615,pastoreEPS}. This Section summarizes the main equations solved within the RAPDENS observer in this novel state estimation algorithm.

The predictive model of the multi-rate EKF is composed of three distinct equations that describe, in a control-oriented approach, the particle balance conservation in terms of plasma electron density and its neutral main species.
\small{
\begin{flalign}
&\dfrac{\partial}{\partial t}(n_{e}V^{'})-\dfrac{\partial}{\partial \rho}\left[ V^{'}\left(G_{1}D\dfrac{\partial n_{e}}{\partial \rho}+ G_{0}\nu n_{e}\right) \right]=SV^{'}& \label{eq:density1d_eq} \\
& \frac{d N_{v}}{dt} = \mathrm{\Gamma_{rec}-\Gamma_{iz}+\Gamma|_{\rho=\rho_{e}}+\Gamma_{recy.}+\Gamma_{valve}-\Gamma_{pump}}&\label{eq:vacuum_eq}\\
&\frac{d N_{w}}{dt} = \mathrm{\Gamma_{SOL \rightarrow wall}-\Gamma_{recy.}}& \label{eq:wall_eq}
\end{flalign}
}
with boundary conditions:
\begin{flalign}
&\dfrac{\partial n_{e}}{\partial \rho}(\rho=0,t)=0,  \phantom{..}n_{e}(\rho=\rho_{e},t)=0&\label{eq:BC_ne}
\end{flalign}
\noindent
and the geometrical coefficients:
\begin{flalign}
&V^{'}=\dfrac{\partial V}{\partial \rho}, \phantom{.}\braket{|\nabla \rho|}=\dfrac{\partial \rho}{\partial \psi}\braket{|\nabla \psi|}\label{eq:geo_coeff1}\\
&G_{0}=\braket{|\nabla \rho|}, \phantom{.}G_{1}=\braket{(\nabla \rho)^{2}}=\left(\dfrac{\partial\rho}{\partial\psi}\right)^{2}\braket{(\nabla \psi)^{2}}& \label{eq:geo_coeff2}
\end{flalign}
\noindent

Equation (\ref{eq:density1d_eq}) solves the 1D flux-surface averaged electron density equation, describing the evolution of the electron density profile $n_{e}(\rho,t)$ over the normalized toroidal coordinate $\rho$, defined as $\mathrm{\rho=\sqrt{\dfrac{\Phi}{\Phi_{\mathrm{LCFS}}}}}$, where $\Phi$ is the plasma toroidal flux and $\Phi_{\mathrm{LCFS}}$ its value computed on the LCFS. 

The flux-surface average operator is denoted with the brackets $\braket{...}$ in Equations (\ref{eq:geo_coeff1}) and (\ref{eq:geo_coeff2}). The spatial domain in which Equation (\ref{eq:density1d_eq}) is solved is $\rho\in [0,\rho_e]$, where $\rho_e = 1+\mathrm{\lambda_{\mathrm{SOL}}}$ accounts for the normalized SOL plasma width $\mathrm{\lambda_{\mathrm{SOL}}}$. In this work, a fixed value of $\mathrm{\lambda_{\mathrm{SOL}}}=0.061$ has been heuristically set. The SOL poloidal density profile dynamics is modeled within the domain $\rho\in [1,1+\mathrm{\lambda_{\mathrm{SOL}}}]$, with the extrapolation of transport coefficient and geometrical quantities defined up to the LCFS. The applied boundary conditions in the solution of the diffusion-advection transport equation for $n_{e}$ are shown in Equation (\ref{eq:BC_ne}), where homogeneous Neumann and Dirichlet boundary conditions have
been applied respectively at $\rho=0$ and $\rho=\rho_{e}$. A further extension and generalization of the boundary conditions, numerically implemented in the solution of the 1D transport equation, have been recently shown in \cite{Kropackova2025}, where Neumann, and Dirichlet inhomogeneous boundary conditions for $n_{e}(\rho)$ have been tested on simulations with AUG real-time data, on the restricted spatial domain $\rho\in[0,1]$. A separate discussion on the transport coefficients $D$ and $\nu$, representing the electron diffusivity and pinch velocity profiles of Equation (\ref{eq:density1d_eq}), is addressed in Section \ref{section:estimation_nu_TS}.

In this work, the experimental results on TCV are obtained under the homogeneous boundary conditions shown in Equation (\ref{eq:BC_ne}). The possibility of testing the multi-rate $n_{e}$ observer with the generalized boundary conditions developed in \cite{Kropackova2025} is explored in \ref{section:Appendix_inhomog_BC}, where an offline simulation demonstrates improved edge profile reconstruction. 

The evolution of the neutrals in the vacuum chamber and wall is described by two 0D ODEs in Equations (\ref{eq:vacuum_eq}) and (\ref{eq:wall_eq}). 
The coupling between the 1D equation for the evolution of $n_{e}=n_{e}(\rho)$ and the 0D equations for $N_v$ and $N_w$ occurs in the source term $\mathrm{S=S(\rho,t)}$:
\small{
\begin{flalign}
&\mathrm{S = S_{iz}-S_{rec}-S_{SOL\rightarrow wall}+S_{NBI}+S_{pellets}}\label{eq:source_term}
\end{flalign}
The non-linear interaction of the neutrals $N_v$ in the vacuum chamber and the plasma edge is retained in the ionization and recombination terms $\mathrm{S_{iz}}$ and $\mathrm{S_{rec}}$; meanwhile, the parallel transport in the SOL region is modeled with the $\mathrm{S_{SOL\rightarrow wall}}$ term. 

For a complete description of each term present in Equations (\ref{eq:vacuum_eq}), (\ref{eq:wall_eq}) and (\ref{eq:source_term}), the reader is referred to \cite{Blanken2018}, where the details of the RAPDENS predictive model and a first formulation of the EKF algorithm based on real time FIR data on TCV and AUG are presented. Moreover, no in-depth analysis of the results concerning the estimation of neutrals contained in the divertor region or wall is reported in this work, due to limited/inexistent diagnostic information on the latter and no integration of neutral measurements in the EKF procedure for the TCV version of the code.  

The 1D transport equation is numerically discretized in space with the finite elements method based on cubic B-splines, as shown in \cite{Blanken2018}, and in time with the Crank-Nicolson scheme \cite{CrankNicolson1947}, tuned for this work in fully-implicit mode. The system of Equations (\ref{eq:density1d_eq}) - (\ref{eq:source_term}) is then cast in the following form:

\begin{flalign}   &\hat{\textbf{x}}_{k|k-1}=\textbf{f}_{d}(\textbf{p}_{k-1},\hat{\textbf{x}}_{k-1|k-1}) + B_{d} \textbf{u} _{k-1}& \label{eq:predictive_eq}
\end{flalign}
The tokamak particle state ${\textbf{x}_k}$, at the time instant $t_{k}$, is composed of the values of spline coefficients that approximate the $n_{e}$ profile and the neutrals inventories described in Equations (\ref{eq:vacuum_eq}) and (\ref{eq:wall_eq}).

To estimate the value of the unknown state ${\textbf{x}_k}$, a two-step approach based on a prediction and correction of the state is carried out with the EKF algorithm. The predicted state $\hat{\textbf{x}}_{k|k-1}$ is provided by Equation (\ref{eq:predictive_eq}), based on the estimated state of the previous iteration in time $\hat{\textbf{x}}_{k-1|k-1}$, on the parameter array $\textbf{p}_{k}$ containing the geometrical quantities reported in Equations (\ref{eq:geo_coeff1}) and (\ref{eq:geo_coeff2}), a boolean description of the plasma state (limited/diverted, L/H-mode), plasma current Ip and volume Vp. The input array $\mathrm{\textbf{u}_{k} =[\Gamma_{valve,k},\Gamma_{NBI,k},\Gamma_{pellet,k}]^{T}}$ contains the particle fluxes provided by the actuators to the plasma and vacuum inventories.
In this work, in line with the model presented in \cite{PASTORE2023113615}, only $\mathrm{u_k=\Gamma_{valve,k}}$ has been employed.
The array $\textbf{f}_{d}$ represents the non-linear term of the state evolution equation, discretized in time. The matrix $B_d$ incorporates, instead, the effect of the actuators on the predicted state with a linear model in a time-discrete representation.

A key step in the design of the multi-rate density observer resides in the formulation of the forward diagnostics model, which maps the plasma particle state ${\textbf{x}\in\mathbb{R}^{n_{x}}}$ onto the measurements space $\textbf{h}\in\mathbb{R}^{n_{h}}$ in the following way:

\begin{flalign}
    &\hat{\textbf{h}}_{k|k-1}    = H_{k}\hat{\textbf{x}}_{k|k-1},\ \ \  H_{k}=\begin{bmatrix}H_{\mathrm{FIR}_{k}} \\ H_{\mathrm{TS}_{k}} \end{bmatrix}&\label{eq:synth_meas_eq}
\end{flalign}

The equilibrium information, provided by the real-time magnetic equilibrium solver, describes the mapping between each diagnostic measurement location in the $(R,Z)$ plane and the corresponding coordinate in the $\rho$ grid, and it is expressed via the linear relationship\footnote{A linear relation between the density profile $n_{e}$ and the FIR and TS measurements can be established since the projection of the profile in the (R, Z) plane is linear with the computation of the spline coefficients at the locations of $\rho_{k}=\rho_{k}(R,Z)$. The sampling/integrals of the density profile mapped onto the FIR and TS chords in the poloidal plane are also a linear operation, resulting in Equation (\ref{eq:synth_meas_eq}) being expressed as a matrix multiplication involving the spline coefficients contained in ${\textbf{x}}_{k|k-1}$ and the matrix $H_{k}$.} in Equation (\ref{eq:synth_meas_eq}). Details on the assembly of the $H_k$ matrix, which depends on the availability of RT-TS data information, can be found in \cite{PASTORE2023113615}. Measurement covariance matrices, associated with the FIR and TS diagnostics, express the weight of each diagnostic's channel in the EKF correction step. Assembled as diagonal matrices\footnote{Each channel's measurement is considered statistically independent from the others, with an additive white noise modeled with the normal distribution $\textbf{v}_k \mathtt{\sim}\mathcal{N}(0,R_{k})$.}, the i-th diagnostic channel's measurement covariance is defined as:
\begin{flalign}
    & R_{\mathrm{diag},k}(i,i)=\sigma_{\mathrm{diag},k,i}^{2} &\label{eq:measurement covariance_eq}
\end{flalign}

Where $\sigma_{\mathrm{diag},k,i}$ represents the normalized standard deviation of the signal on the i-th channel at the time $t_k$.

From the information retained in the synthetic diagnostics prediction $\hat{\textbf{h}}_{k|k-1}$ and the actual, normalized, measurements stacked in $\textbf{h}_{k}$, it is possible to compute the innovation residual, defined as:
\begin{flalign}
    &{\hat{\textbf{z}}}_{k}=\textbf{h}_{k}-{\hat{\textbf{h}}}_{k|k-1} &\label{eq:innovation_res}
\end{flalign}
Real-time monitoring of the innovation residual ${\hat{\textbf{z}}}_{k}$ can be leveraged to isolate diagnostic faults and offsets, provided that the synthetic diagnostic model is sufficiently accurate. Examples based on this approach have been implemented in RAPDENS \cite{Blanken2018,Bosman2021} to flag and correct for possible FIR fringe jumps, and in \cite{PASTORE2023113615} to infer and compensate for offsets present between FIR and TS data. The corrected signal, named ${\hat{\textbf{z}}}_{\mathrm{corr},k}$, is used as a basis for the correction step of the EKF. The excluded diagnostics channels, either discarded in real time through the analysis of ${\hat{\textbf{z}}}_{k}$ or manually deselected by the user before the shot, are removed from a selection matrix $\mathbb{S}_{k}$ that activates the diagnostic channels in the correction step of the observer. The selection matrix $\mathbb{S}_{k}$ is built from an identity matrix of size $n_h \times n_h $, where $n_h$ is the total number of diagnostic channels, from which the deactivated channels' corresponding rows are removed. In this work, the selection matrix is designed to discard the FIR channels located on the inboard side of the vacuum vessel, since they are usually more prone to fringe jumps and do not further increase the information on the density profile, and additionally discard TS data present outside the LCFS, not resolved in the density model prediction expressed by Equation (\ref{eq:density1d_eq}).

The model $\textbf{f}_d$ is linearized over the predicted state $\hat{\textbf{x}}_{k|k-1}$, and the associated Jacobian matrix $F_{k}$ is adopted for the computation of the prediction error covariance matrix $P_{k|k-1}$, in Equation (\ref{eq:pred_cov_matrix}) of the EKF. The state covariance matrix $Q_{k}$ here represents the covariance associated with the elements of $\textbf{x}_k$, and it is modeled as additive white noise $\textbf{w}_k \mathtt{\sim}\mathcal{N}(0,Q_{k})$. The values of this matrix, which are tuned for experimental use in real time for the observer, are reported in \ref{section:Appendix_covariance_matrices}. 

The Extended Kalman gain $L_{k}$ is then assembled in Equation (\ref{eq:Kalman_gain_matrix}) as:

\begin{flalign}
&P_{k|k-1}=F_{k-1}P_{k-1|k-1}F_{k-1}^{T}+Q_{k-1}& \label{eq:pred_cov_matrix} \\
&S_{k}=H_{k}P_{k|k-1}H_{k}^{T}+R_{k}& \label{eq:inn_res_cov_matrix} \\
&L_{k}=P_{k|k-1}H_{k}^{T}\mathbb{S}_{k}^{T}(\mathbb{S}_{k}S_{k}\mathbb{S}_{k}^{T})^{-1}& \label{eq:Kalman_gain_matrix} \\
&P_{k|k}=(I-L_{k}\mathbb{S}_{k}H_{k})P_{k|k-1}& \label{eq:posterior_cov_matrix} 
\end{flalign}

The ancillary Equations (\ref{eq:inn_res_cov_matrix}),(\ref{eq:posterior_cov_matrix}) compute the innovation residual covariance matrix $S_{k}$, and the $P_{k|k}$ posterior error covariance matrix, respectively. The selection matrix $\mathbb{S}_{k}$ acts as an index contraction from the total number of diagnostic channels to the restricted set of activated channels at the k-th time step. 

As final step, the estimate of the corrected state $\hat{\textbf{x}}_{k|k}$ is computed as: 
\begin{flalign}
&\hat{\textbf{x}}_{k|k}=\hat{\textbf{x}}_{k|k-1}+L_{k}\mathbb{S}_{k}\hat{\textbf{z}}_{\mathrm{corr},k}& \label{eq:correction_step} 
\end{flalign}

The state $\hat{\textbf{x}}_{k|k}$ incorporates the model prediction $\hat{\textbf{x}}_{k|k-1}$ information and the contribution coming from the measurements $L_{k}\mathbb{S}_{k}\hat{\textbf{z}}_{\mathrm{corr},k}$, based on the corrected innovation residual $\hat{\textbf{z}}_{\mathrm{corr},k}$. The Extended Kalman gain $L_{k}$ acts as a weight between measurements and model prediction, and it can be tuned by shifting the importance of the $Q_k$ model covariance matrix over the $R_k$ measurements covariance matrix \footnote{One can quantify the importance of one covariance matrix over the other by associating, for each matrix, its matrix norm and rescaling the matrices with scalar coefficients. This shifts the weight of the reconstruction over model and diagnostics, and internally in the diagnostic model, among the different diagnostics composing the forward diagnostic model. }.

For a more detailed description of the real-time algorithm, the reader is referred to the \ref{section:Appendix_scheme_observer}.

\subsection{Estimation of the pinch velocity-to-diffusivity ratio's radial profile}\label{section:estimation_nu_TS}

The information contained in the update step, Equation (\ref{eq:correction_step}), whenever updated RT-TS data is available, is used to adapt the predictive model and FIR measurements fed into the multi-rate observer in real time. The $n_{e}$ profile is constrained by local density measurements from the RT-TS system and mapped onto the $\rho$ grid using the magnetic equilibrium solver. To propagate such spatially detailed information to the model, an analytical formulation of the electron pinch velocity-to-diffusivity ratio has been adopted, based on the steady-state expression of the 1D flux-surface-averaged electron density equation. The $n_{e}$ profile information, the profile of the diffusivity transport coefficient $D=D(\rho)$, heuristically tuned for L- and H- mode density profiles (in Figure \ref{fig:diffusion_coeff}), and the geometric coefficients reported in Equations (\ref{eq:geo_coeff1}) and (\ref{eq:geo_coeff2}) are used to extract the relation on $\nu=\nu(\rho)$.

Details on the implementation are shown in \cite{pastoreEPS}. Here, for clarity, the main steps are highlighted.

The steady-state form of Equation (\ref{eq:density1d_eq}) is used:
\begin{flalign}
&-\dfrac{\partial}{\partial \rho}\left[ V^{'}\left(G_{1}D\dfrac{\partial n_{e}}{\partial \rho}+ G_{0}\nu n_{e}\right) \right]=SV^{'}& \label{eq:steady_density1d_eq}
\end{flalign}

Integrating over $\rho$ on both sides of Equation (\ref{eq:steady_density1d_eq}), applying the homogeneous boundary conditions (\ref{eq:BC_ne}) and isolating the ratio $\nu/D$ leads to:
\begin{flalign}
&\phantom{-}\dfrac{\nu_{\mathrm{TS}}}{D}= -\dfrac{G_{1}}{G_{0}} \dfrac{1}{n_{e,\mathrm{TS}}}\dfrac{\partial n_{e,\mathrm{TS}}}{\partial \rho}-\dfrac{\hat{S}}{DG_{0}n_{e,\mathrm{TS}}},\ \ \hat{S}=\dfrac{1}{V{'}}\int_{0}^{\rho} S V^{'}d\rho & \label{eq:nu_eq}
\end{flalign}

The subscript TS indicates here that the updated electron density profile $n_{e,\mathrm{TS}}$ state using RT-TS data is adopted in the estimation of the pinch velocity term $\nu_{\mathrm{TS}}$.

The estimated $\nu_{\mathrm{TS}}/D$ ratio is composed of two additive terms. The first one scales as the inverse logarithmic scale length $\dfrac{1}{n_{e,\mathrm{TS}}}\dfrac{\partial n_{e,\mathrm{TS}}}{\partial \rho}$. Note that the exact values of $D$ or $\nu$ are not significant for stationary profiles, since only $\nu/D$ counts in such a limit. The second term incorporates instead the dependence of the density profile on the source term $\hat{S}$, as described in Equation (\ref{eq:source_term}). 
In this study, the source term $\hat{S}$ in Equation~(\ref{eq:nu_eq}) is neglected for simplicity. This choice is motivated by the lack of accurate information on the spatial distribution of neutrals that penetrate the plasma from the separatrix inward towards the core, and the unavailability of real-time NBI deposition profiles. Although the contribution of gas fueling is relatively small, it produces a measurable effect near the plasma edge, as the primary ionization source, heuristically modeled, is localized in the region $\rho \ge 0.60$. The evaluation of the NBI fueling rate, which is predominantly concentrated in the plasma core, can be performed using dedicated codes, such as RABBIT \cite{Weiland2018}, or with an ad hoc model for the NBI deposition function. A natural extension of this work is to estimate the source within the EKF framework using disturbances, as shown in \cite{VanMulders2025}.

The implementation in the observation loop is set up by estimating the electron pinch velocity $\nu_{TS}$ as a function of the logarithmic electron density gradient and the diffusivity profile $D$ (Figure \ref{fig:diffusion_coeff}), thus updating the ratio in a real-time compatible manner with newly available TS data at 60 Hz. In addition, an inversely proportional scaling of the diffusivity $D$ with respect to the plasma current is applied to provide a time-varying dependence of the diffusivity during plasma current ramp-up or ramp-down. Details on this can be found in \cite{Blanken2018}.

\begin{figure}[h!]
\centering
  \includegraphics[width=0.55\textwidth]{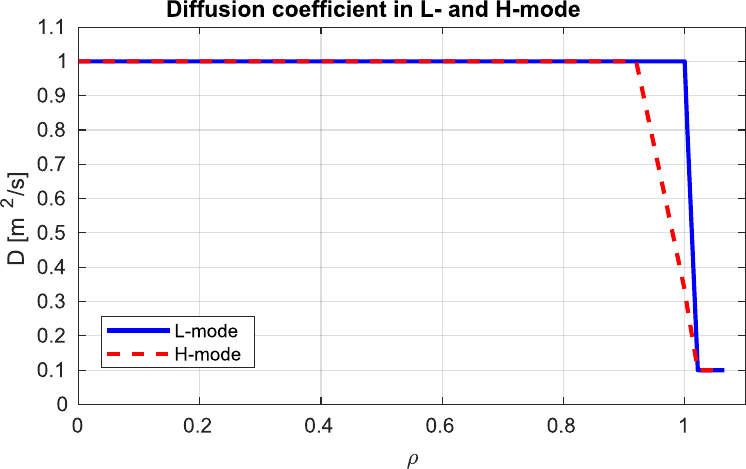}
  \centering
  \caption{ The diffusion transport coefficient $D \ [m^{2}/s]$ heuristically tuned for L- and H-mode density profiles. }
  \centering
  \label{fig:diffusion_coeff}
\end{figure}

\subsection{Numerical validation of the estimation of $\nu/D$ transport coefficient ratio in RAPDENS predictive model}\label{section:validation_nu_TS}

To test the methodology presented in Section \ref{section:estimation_nu_TS}, a simulation is carried out with the RAPDENS predictive model using a synthetic density profile, to verify in particular the impact of the simplifying hypothesis made on the prediction performance.

A (target) analytical electron density profile is defined beforehand, shown in dashed red in Figure \ref{fig:timeslices_synth_nu_D}, and the corresponding $\bar{\nu}$ transport coefficient is computed from the analytical profile's logarithmic gradient and a fixed diffusivity coefficient $\bar{D}$ (the L-mode coefficient shown in Figure \ref{fig:diffusion_coeff}, with no additional plasma current rescaling), using Equation (\ref{eq:nu_eq}) and neglecting the additional source term $\hat{S}$. The initial density profile for the predictive simulation which evolves towards a steady state solution is shown in Figure \ref{fig:timeslices_synth_nu_D}a in a solid black line. In $\approx100$ ms, the profile reaches a steady state solution, close to the target analytical profile. The time evolution of the predicted profile is shown in Figure \ref{fig:timetraces_synth_nu_D}, with a fast component of the time response decaying in $\approx100$ ms and a slow component, in the order of $\approx500$ ms, provided by the wall recycling term. A relative error of 2\% between the target profile and the predicted profile is present, deemed acceptable compared to the accuracy of the TS measurements.

\begin{figure}[h!]
\centering
  \includegraphics[width=0.8\textwidth]{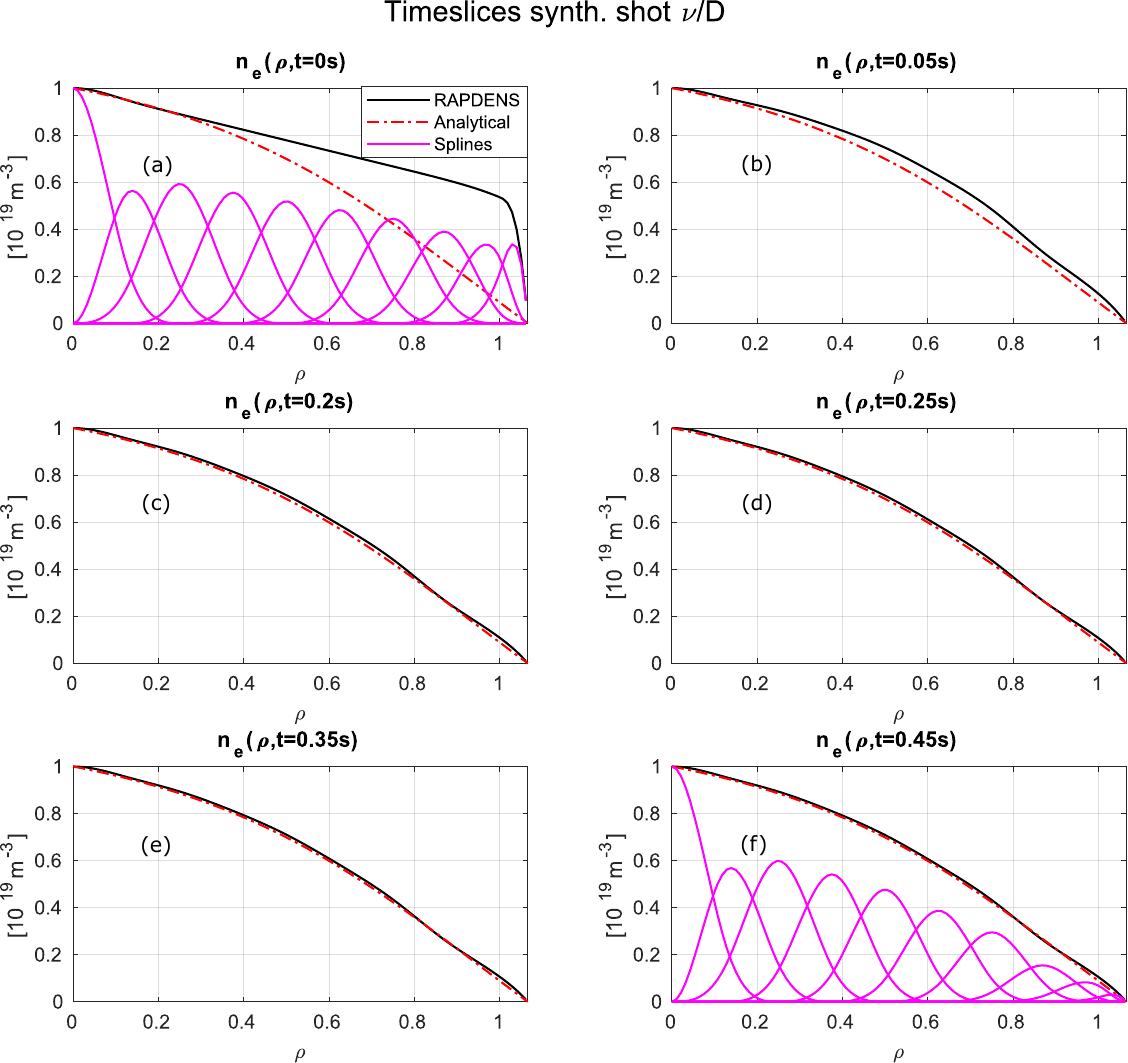}
  \centering
  \caption{ Timeslices of the RAPDENS predictive simulation using an analytical target profile (in dashed red) to provide the $\nu/D$ ratio adopted in the simulation. RAPDENS profiles, in black and spline coefficients, in magenta, are shown. The density profile evolves from the initial condition, in Figure \ref{fig:timeslices_synth_nu_D}a, to the target profile in dashed red.}
  \centering
  \label{fig:timeslices_synth_nu_D}
\end{figure}

\begin{figure}[h!]
\centering
  \includegraphics[width=0.8\textwidth]{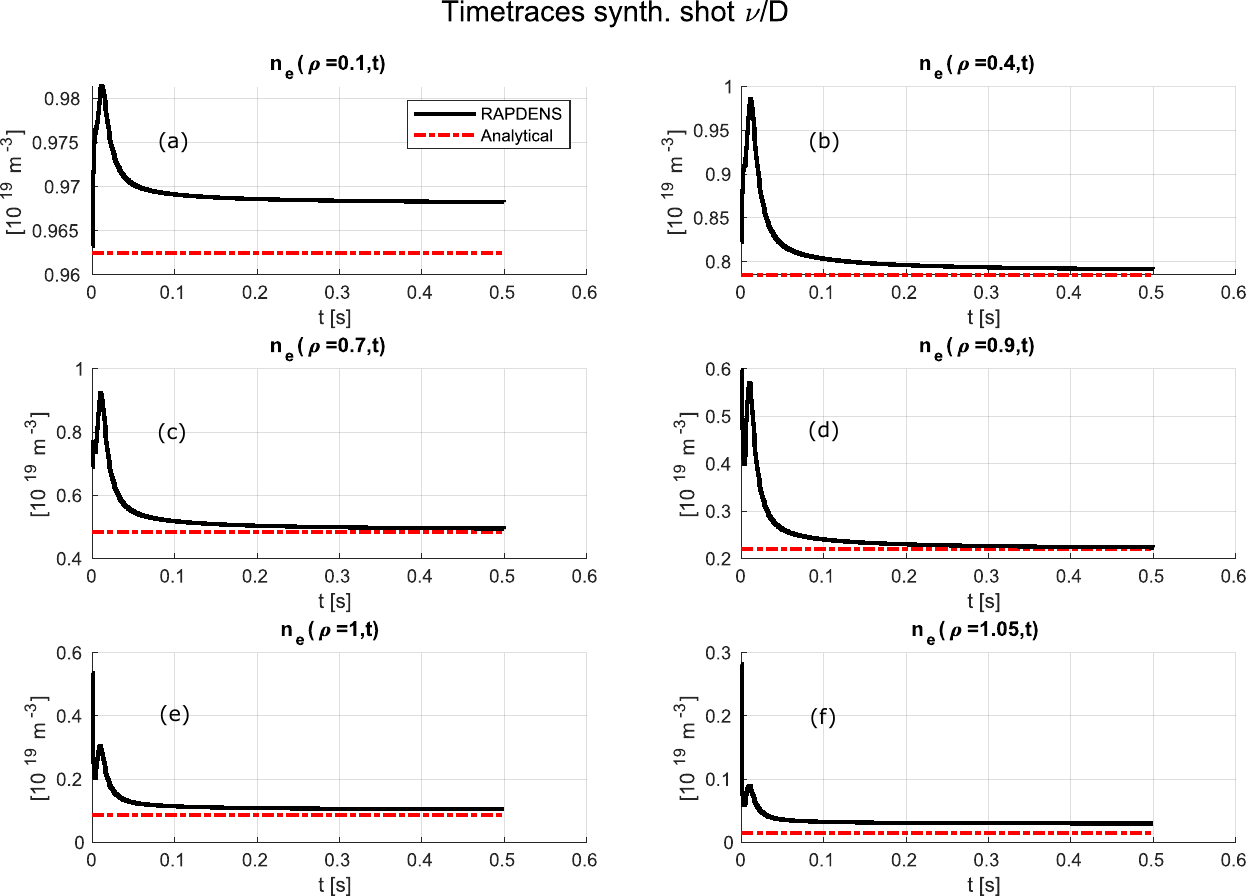}
  \centering
  \caption{Timetraces of the RAPDENS predictive simulation using an analytical target profile (in dashed red) to provide the $\nu/D$ ratio adopted in the simulation. RAPDENS timetraces, at different radial coordinates, are reported in black.}
  \centering
  \label{fig:timetraces_synth_nu_D}
\end{figure}

An example of the real-time estimation of the electron pinch velocity-to-diffusion ratio $\nu/D$ for TCV shot \#82913 is reported in the experimental Section \ref{section: ne control below cutoff}, in Figures \ref{fig:nu 82913 timetraces} and \ref{fig:nu 82913 timeslices}, where the adaptive estimation of the transport coefficient at a frequency equal to the TS system ensures accurate prediction of the density profile in the limited, diverted and ECH phases of the shot. The impact of incorporating the estimated transport coefficient ratio in the spatial resolution of the electron density profile is then explored in Section \ref{section:impact_nu_D_82913}, where a comparison between a fixed transport coefficient ratio and the adoption of the time adaptive one is evaluated.

\section{Control scheme deployed on SCD}\label{section:Control scheme SCD}
In this Section, the main algorithms involved in the novel density observation and control loop are summarized. The integration of these algorithms is carried out within the \textit{Système de Contrôle Distribué} (SCD) \cite{GALPERTI2024114640}, TCV digital PCS. An overview of the details of the processed signals, specifically in the real-time data treatment of digitally filtered FIR data, and sampling time of each signal is presented in \cite{pastoreEPS}. Here, the main signals are briefly presented for completeness, as shown in Figure \ref{fig:SCD_scheme}. In the remaining part of this Section, the different blocks that compose the SCD scheme are described. Detailed information related to the structure of the RAPDENS observer block can be found in \ref{section:Appendix_scheme_observer}.

The real-time compatible diagnostics employed in the multi-rate Extended Kalman Filter involve local and integral density measurement signals of TS and FIR diagnostics, which can be used to constrain the density profile to be observed. A graphical representation of the spatial distribution of the measurement chords for these two diagnostics is represented in Figure \ref{fig:liuqe_TS_FIR}, along with a poloidal section of the TCV tokamak. RT-TS data are produced with 117 polychromators, resolving $T_e$ and $n_e$ data spatially on the corresponding observed scattering volumes distributed along a vertical line in the poloidal plane. The signals are provided by 3 distinct Nd:YAG lasers, individually capable of firing at a frequency of 20 Hz, for a total system frequency of 60 Hz. RT-FIR diagnostic, collecting the data of the 14 vertical FIR chords, provides integral values of the density present along the lines of sight, which are radially distributed in the poloidal plane. These signals, produced at a frequency of 200 kHz, are digitally filtered to 10 kHz to avoid propagation of fringe jumps to the different algorithms on SCD that employ these signals. The aforementioned filtering method increases the reliability of the control system, but does not ensure perfect rejection of fringe jumps, especially in the case of small signal-to-noise ratio (SNR); therefore, a model-based approach for fringe jumps filtering based on the RAPDENS EKF is adopted. The details are further discussed in \ref{section:Appendix_scheme_observer}, and an example of real-time, multiple fringe jump filtering of two corrupted channels is shown in Section \ref{section: ne edge control H-mode}, Figure \ref{fig:82876_high_performance}. Currently, RT error bars are available on SCD for TS and FIR, but due to bandwidth limitations on the framework, they are not propagated to other algorithms. Fixed standard deviations $\sigma_{\mathrm{TS}}$ and $\sigma_{\mathrm{FIR}}$ are therefore introduced in the EKF procedure to assign a given weight to each diagnostic measurement. Their values are reported in \ref{section:Appendix_covariance_matrices}. FIR chords close to the plasma edge are further penalized in the real-time implementation, increasing the relative covariance $R_{\mathrm{FIR},k}$ between the most external chords intersecting the plasma and the interior ones\footnote{A scaling factor of $10^{10}$ is applied to the covariance matrix entry  $R_{\mathrm{FIR},k}(i,i)$ each time the i-th FIR chord results in a total contribution to the FIR forward model $H_{\mathrm{FIR}_k}$ less than a set threshold. This condition is usually verified for FIR chords intersecting only a small fraction of the plasma edge.}, to take into account the errors in the reconstruction of the plasma LCFS provided by the magnetic equilibrium solver. 
 
Plasma state confinement (L- or H-mode) is computed with the LDH (low, dithering, and high confinement) state detector \cite{Marceca} observer, which employs a RT machine learning state detection algorithm based on ConvLSTM \cite{Shi2015}. This algorithm uses photodiode data (PD), FIR, diamagnetic loop (DML), and $I_p$ to provide a boolean signal at 1 kHz to the predictive model of RAPDENS. The boolean is employed to select transport coefficients that best suit the logarithmic scale length and particle confinement time of $n_{e}$ in L- and H-mode, e.g., as shown in Figure \ref{fig:diffusion_coeff} with the diffusion coefficient profiles tuned for these plasma confinement states.

RT-LIUQE (1 kHz) \cite{Moret2015} provides magnetic equilibrium quantities, such as poloidal flux map $\psi(R,Z)$, toroidal flux $\Phi(\psi)$, and flux-surface-averaged quantities reported in Equations (\ref{eq:geo_coeff1}) and (\ref{eq:geo_coeff2}). These quantities are used to compute in real time the required mapping between the spatial location in the poloidal plane of the TS and FIR measurements \cite{PASTORE2023113615} and the normalized toroidal coordinate $\rho$, as well as to inform and update the RAPDENS predictive model. The gas valve flow readback, in particles per second, is also provided to the RAPDENS fueling module.

\begin{figure}[ht!]
\centering
  \includegraphics[width=0.75\textwidth]{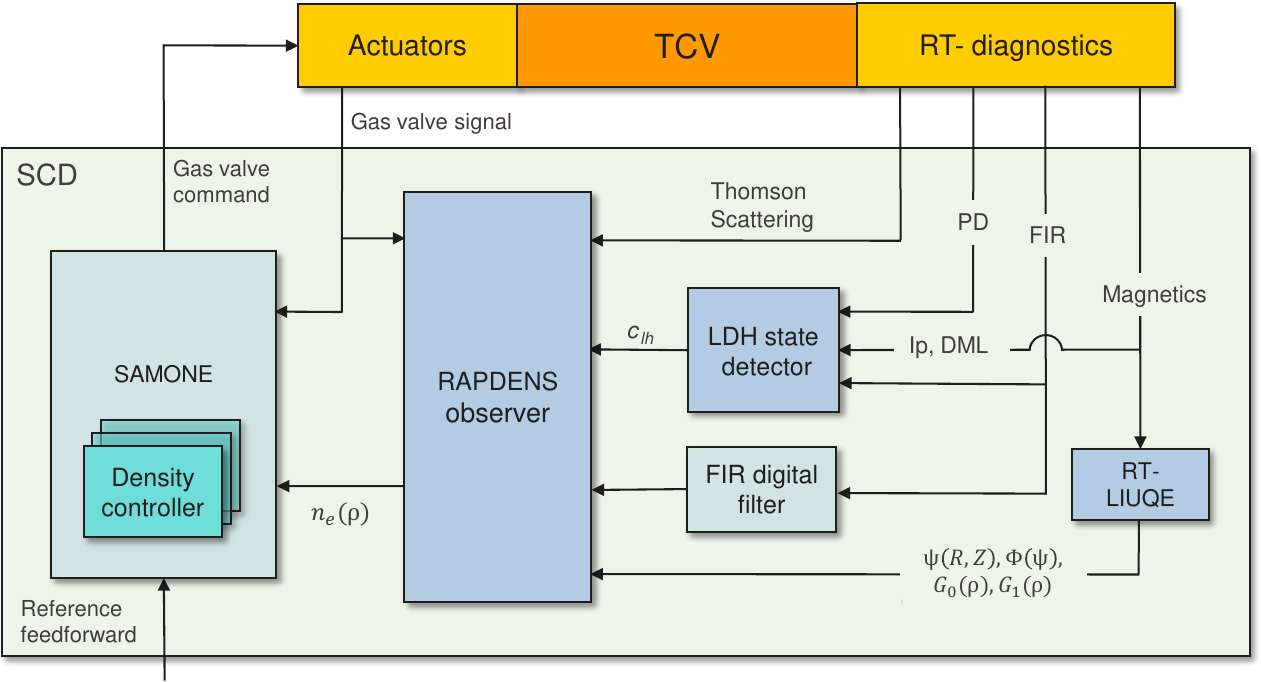}
  \centering
  \caption{\footnotesize{Schematic of the implementation of the electron density observer inside SCD.}}
  \centering
  \label{fig:SCD_scheme}
\end{figure}

The resulting $n_e$ profile output of RAPDENS, computed in less than 1 ms, and its derived quantities, such as the line-integrated and averaged density within the LCFS, are propagated to other algorithms deployed on SCD through a plasma state bus \cite{GALPERTI2024114640}, where different algorithms populate the most significant plasma quantities that describe its state.

The plasma state bus is then transmitted to SAMONE, the Supervisory Actuator Manager and handler of Off-Normal Events \cite{Vu_2021}, used in this work to program the feedback control of the electron density profile or a derived density quantity. Different tasks can be programmed within SAMONE, with the possibility to allocate one or multiple actuators for a single control task. For the work presented in the following Sections, SAMONE tasks for density and beta control have been programmed with fixed time windows. These tasks adopt PI anti-windup controllers, with an additional feed-forward term, for gas valve flux regulation. The controller gains have been tuned heuristically using the model, simulating the actuators' response with a different set of gains using stored real-time signals from previous shots as simulation input. Shot-to-shot adjustment of the gains and feedforward gas trace has also been carried out, in complement to the previous approach. Optimization of the density controller's gains could also be carried out with the control-oriented model within RAPDENS, as shown in \cite{weldon_fueling_2026}. At the current state, the possibility to either perform feedback control of the line-averaged electron density, or the SISO control of a local point of the $n_{e}$ profile for a given $\mathrm{\rho_{target}}$, can be carried out with the digital control scheme depicted in Figure \ref{fig:SCD_scheme}.  


In the following Sections, some examples of the electron density control leveraging the multi-rate state observer are showcased to demonstrate the readiness of our advanced density control framework and understanding for next step device requirements. The application spans from line-average density control in ohmic plasmas with strong impurity-seeding to local density control of the central density below cutoff in ECH L-mode plasmas. Edge density control of high-density, high-performance, H-mode plasmas is also investigated, with simultaneous control of the plasma $\beta_{\mathrm{tor}}$ with NBI.
\section{Control of line-averaged electron density in support of detachment studies}\label{section:Detachment}

The access to divertor detachment \cite{leonard_plasma_2018} can be regulated by the power crossing the separatrix, the upstream plasma conditions at the separatrix, the adoption of different divertor geometries \cite{theiler_results_2017,verhaegh_divertor_2025} (e.g. varying the divertor leg length \cite{reimerdes_tcv_2017}, second-order nulls \cite{Gorno2023}, intercepting the divertor leg with an additonal X-point \cite{Lee2025_XPTR}, or changing the poloidal and total flux expansion \cite{Carpita2024,verhaegh_divertor_2025}), or injection of low-Z impurities \cite{fevrier_nitrogen-seeded_2020,Kallenbach2018}. Upstream SOL conditions are set by the power crossing the separatrix and the total plasma pressure at the separatrix  \cite{Stangeby2000}. For practical reasons, the control of plasma pressure at the separatrix is indirectly carried out by controlling the electron density at the separatrix, $n_{e,\mathrm{sep}}$, since the temperature is mainly set up by the power source/sink terms within the LCFS for core performance purposes. A proportionality factor between the line-integrated density and $n_{e,\mathrm{sep}}$ can be established, provided that the density profile peaking factor does not change in the different sections of the shot \footnote{The effect of external heating actuators, or injection of impurities on the underlying transport mechanisms that impact the $n_e$ profile peaking will be studied in more detail in Section \ref{section: ne control below cutoff}.} and that the density outside the LCFS does not add a too large offset to the measured signal. The control of $\mathrm{NEL}$, the line-averaged electron density derived from the central FIR channel signal, can then be used as a proxy to control the upstream density conditions. For the TCV case, $\mathrm{NEL}$ is indirectly controlled using the vertical FIR \#6 line-integral density signal $y_{\mathrm{FIR},6}$, whose line of sight coincides with the TS line of sight, see Figure \ref{fig:liuqe_TS_FIR}. The post-shot computation of NEL is then provided using the LIUQE reconstruction of the intersecting length $l_{\mathrm{FIR},6}$ of FIR \#6 with the plasma LCFS in the poloidal plane. The line-averaged electron density is ultimately computed as $\mathrm{NEL}=y_{\mathrm{FIR},6}/l_{\mathrm{FIR},6}$.

The aforementioned scheme for controlling the upstream density has a major drawback: the pickup of a non-negligible SOL density in the divertor region, especially in high-density plasma regimes, in detached plasmas, or in the presence of enhanced neutral recycling from the wall. The line-averaged density computed with the vertical interferometer can be decomposed into two contributions,  $\mathrm{NEL}=\mathrm{NEL}_{\mathrm{LCFS}}+\mathrm{NEL}_{\mathrm{SOL}}$, where $\mathrm{NEL}_{\mathrm{LCFS}}$ is the line-averaged electron density reconstructed within the LCFS, and $\mathrm{NEL}_{\mathrm{SOL}}$ is the remaining contribution in the SOL.

\begin{itemize}
    \item $\mathrm{NEL}_{\mathrm{LCFS}}$ is computed using the density profile estimate remapped on the magnetic equilibrium, assuming a poloidally symmetrical distribution of the density profile for each flux surface contained within the separatrix, and integrated along the central FIR \#6 line of sight which crosses the plasma within the LCFS, $\mathrm{NEL}_{\mathrm{LCFS}}=y_{\mathrm{LCFS}}/l_{\mathrm{FIR},6}$ .
    \item $\mathrm{NEL}_{\mathrm{SOL}}$ is instead simply determined by subtraction of the two previously defined quantities $\mathrm{NEL}_{\mathrm{SOL}}=\mathrm{NEL}-\mathrm{NEL}_{\mathrm{LCFS}}$, providing the integrated value in the SOL $y_{\mathrm{SOL}}=\mathrm{NEL}_{\mathrm{SOL}} l_{\mathrm{FIR},6}$.
\end{itemize}

The measured line-integrated signal $y_{\mathrm{FIR},6}$ used for standard density feedback control, in case of non-negligible contributions of $\mathrm{NEL}_{\mathrm{SOL}}$, no longer has a linear relation with the density integrated within the LCFS. The signal affected by the density pickup in the SOL propagates to the controller, which in turn provides a gas flux command that modifies the density upstream conditions in an undesirable way, usually leading to a $\mathrm{NEL}_{\mathrm{LCFS}}$ drifting w.r.t. the target over time, as the $\mathrm{NEL}_{\mathrm{SOL}}$ increases (as shown in detail in Figures \ref{fig:RAPDENS-based control of NEL,LCFS}, \ref{fig:RAPDENS-based control of NEL,FIR}). 
  
Different physical mechanisms affect the density present in the SOL. Ionization reactions of the neutrals injected from the gas valve in the divertor region increase the SOL density. Recycling of the neutral particles from the wall, which is enhanced as additional power crosses the separatrix, and the injection of extrinsic impurities to favor plasma detachment can also lead to a progressive increase in the plasma SOL density, especially in deeply detached plasmas and MARFE \cite{Lipschultz1984} formation.

This study employs the multi-rate electron density observer to provide more precise estimates of $\mathrm{NEL}_{\mathrm{LCFS}}$. RT-LIUQE computes in real time the magnetic equilibrium, useful to track the plasma domain and recompute at each time step the mapping between density measurements and plasma density state (as shown in Equation (\ref{eq:synth_meas_eq})), and RT-TS data are leveraged to constrain the electron density profile while correcting for interferometer offsets due to the pick-up of the SOL density. This renders the framework to remain valid even when the shape changes during a shot. $\mathrm{NEL}_{\mathrm{LCFS}}$, reconstructed by the density observer, can then be leveraged to provide a reference signal to the density controller, which is unaffected by the contribution coming from the SOL density pickup. This density control scheme leads to a decoupling between upstream and SOL conditions, providing an experimental framework where SOL conditions and/or divertor magnetic geometry can be varied during the shot or in different shots, while keeping the density upstream conditions controlled independently, as will be necessary in fusion power plants.

\subsection{Control of $\mathrm{NEL}_{\mathrm{LCFS}}$ for ohmic, nitrogen-seeded plasmas: scan of the outer divertor leg length}\label{section:control_NEL_LCFS}
A detachment study conducted using the traditional density control scheme resulted in unsatisfactory density control within the LCFS. The scenario, in fact, is prone to a time-varying pick up of the density in the divertor region, leading to a non-constant value of $\mathrm{NEL}_{\mathrm{LCFS}}$ as the shot evolves. Such a study has been used as a testing ground for the novel density control scheme. The detachment experiment consists of exploring the effect of the outer divertor leg length on the detachment onset in three separate shots. An ohmic scenario at 300 kA has been selected, with $N_2$ seeding prescribed as a feedforward ramp ($\mathrm{t=[0.900 \ s - 1.80 \ s]}$, $\mathrm{GV}_{N_2}=\mathrm{[0\ molec./s - 6.25 \cdot 10^{20} \ molec./s]}$) and $D_2$ feedback density fueling. The timetraces are shown in Figure \ref{fig:gas_leg_scan}. The shaping flexibility of TCV in generating such equilibria and its highly elongated vacuum vessel has been leveraged to position the plasma at different vertical positions, as is visible in Figure \ref{fig:liuqe_TS_FIR}. In this experimental comparison, it is necessary to maintain controlled upstream conditions for a valid comparison of the downstream plasma behavior, to isolate the effect of the varying leg-length on plasma detachment.

\begin{figure}[h!]
\centering
  \includegraphics[width=0.5\textwidth]{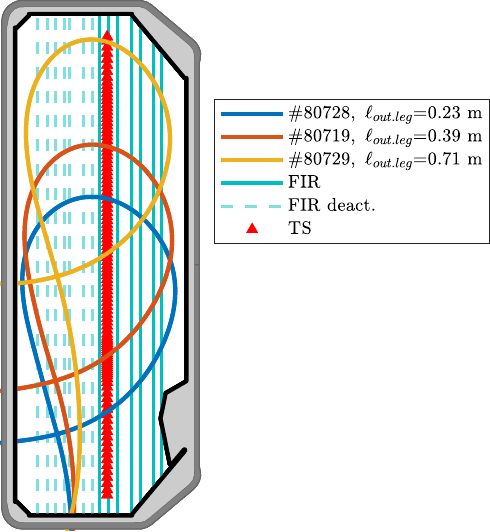}
  \centering
  \caption{Poloidal section of TCV. The LCFS configurations of the shots \#80728 (short leg, in blue), \#80719 (medium leg, in orange), and \#80729 (long leg, in yellow) are reported. The FIR system is represented in cyan, with 14 vertical lasers crossing the vacuum chamber. 
  The numbering order of each channel starts with FIR \#1 on the LFS, moving inwards towards FIR \#14 on the HFS.
  The channels \#1 to \#7, represented with solid lines, are employed in the RAPDENS EKF reconstruction. The remaining set of FIR channels, from \#8 to \#14, is represented with dashed lines. Thomson scattering system's 117 closely packed volume measurement points at R=0.900 m, aligned with FIR laser \#6 at R=0.903 m, are depicted with red triangle markers.
    }
  \centering
  \label{fig:liuqe_TS_FIR}
\end{figure}

A set of shots with the traditional control scheme is reported in Figures \ref{fig:RAPDENS-based control of NEL,LCFS}a and \ref{fig:RAPDENS-based control of NEL,FIR}a. The color code reflects the one used in Figure \ref{fig:liuqe_TS_FIR}. The black dashed line represents the $\mathrm{NEL}_{\mathrm{LCFS}}$ target aimed throughout the experiment. $\mathrm{NEL}_{\mathrm{FIR}}$ chord $\#6$  is used with the traditional control scheme. In Figure \ref{fig:RAPDENS-based control of NEL,FIR}a, the $\mathrm{NEL}_{\mathrm{FIR}}$ overlaps in the short (in blue) and medium-legged (in orange) shots, indicating that the control task is reproduced successfully. A manual increase in the feedforward $D_2$ gas trace is instead applied in the long-legged case (in yellow), to compensate for the higher pickup of SOL density, see Figure \ref{fig:gas_leg_scan}a. $\mathrm{NEL}_{\mathrm{LCFS}}$, in Figure \ref{fig:RAPDENS-based control of NEL,LCFS}a, which is computed using the post-shot TS profile, remains near the target from  $0.750 \ s$ to $1.00 \ s$  only for the short- and medium-legged configurations. As the shot continues with the nitrogen ramp injection, a departure from the target density is noticeable due to an increase of $\mathrm{NEL}_{\mathrm{SOL}}$  favored by the impurity seeding, from $t=0.900 \ s$. The $\mathrm{NEL}_{\mathrm{LCFS}}$ of the long-legged case remains close to the target only up to $t=0.800 \ s$, due to the higher contribution of the FIR signal pickup in the divertor region for this configuration. A strong decrease of $\mathrm{NEL}_{\mathrm{LCFS}}$ in the long-legged case can be noticed, leading to an overall mismatch of $\approx30\%$ from the target density at the end of the shot.

This set of shots is then repeated, employing the RAPDENS-SAMONE density control scheme for direct control of $\mathrm{NEL}_{\mathrm{LCFS}}$ to perform the upstream density control task. The control task starts at $\mathrm{t\geq0.55 \ s}$ for the three cases, visible in Figure \ref{fig:gas_leg_scan}b, with a user-defined saturation of the $D_2$ fueling at $\mathrm{ 7.50 \cdot 10^{20} molec./s}$ to avoid excessive injection of neutrals in this section of the shot. The results of the upstream density control are reported on Figures \ref{fig:RAPDENS-based control of NEL,LCFS}b and \ref{fig:RAPDENS-based control of NEL,FIR}b. The control of $\mathrm{NEL}_{\mathrm{LCFS}}$ has been successfully achieved, as shown in Figure \ref{fig:RAPDENS-based control of NEL,LCFS}b, as the three shots exhibit similar control performances close to the target density, reduced from $5.9$ to $\mathrm{5.0 \cdot 10^{19} m^{-3}}$ for exhaust scenario considerations.

A consistent offset of $\approx 10\%$ from the target is present at the end of the three shots with the RAPDENS-based approach, due to the difference in the computation of $\mathrm{NEL}$ in the real-time framework, shown in \ref{fig:RAPDENS-based control of NEL,FIR}b in darker shades, compared with the offline computation of $\mathrm{NEL}$ in Figure \ref{fig:RAPDENS-based control of NEL,LCFS}b. The real-time computation of the intersection of the FIR \#6 with the plasma employed a simplified formula $l_{\mathrm{FIR},6} = 2 ka$, where $k$ and $a$ are elongation and minor radius of the plasma, provided by RT-LIUQE. The offline one is based instead on the LIUQE equilibrium reconstruction, computing the intersection of the FIR central chord within the plasma LCFS with an offline numerical algorithm.
This problem has been recently resolved by using the real-time computation of the FIR length crossing the plasma domain, computed directly from the synthetic diagnostic model of the RAPDENS observer, as shown in Equation (\ref{eq:synth_meas_eq}).

As a dual case to what was described beforehand in the traditionally-controlled result section, a linear, ramp-like trend in the time evolution of $\mathrm{NEL}_{\mathrm{FIR}}$ can be noticed in Figure \ref{fig:RAPDENS-based control of NEL,FIR}b, which is further enhanced in the long-legged case in yellow. An increase of $\approx 40\%$ of $\mathrm{NEL}_{\mathrm{FIR}}$ results from the $N_2$ impurity seeding, as shown in Figure \ref{fig:gas_leg_scan}b in scales of gray, while the $\mathrm{NEL}_{\mathrm{LCFS}}$ is kept constant and close to the desired density target. The performance is maintained even in the undesired case of saturation of the impurity seeding gas flow at sightly different levels, due to unexpected reduction in the upstream neutral pressure of the $N_2$ gas line in between the shots. 



\begin{figure}[h!]
\centering
  \includegraphics[width=0.95\textwidth]{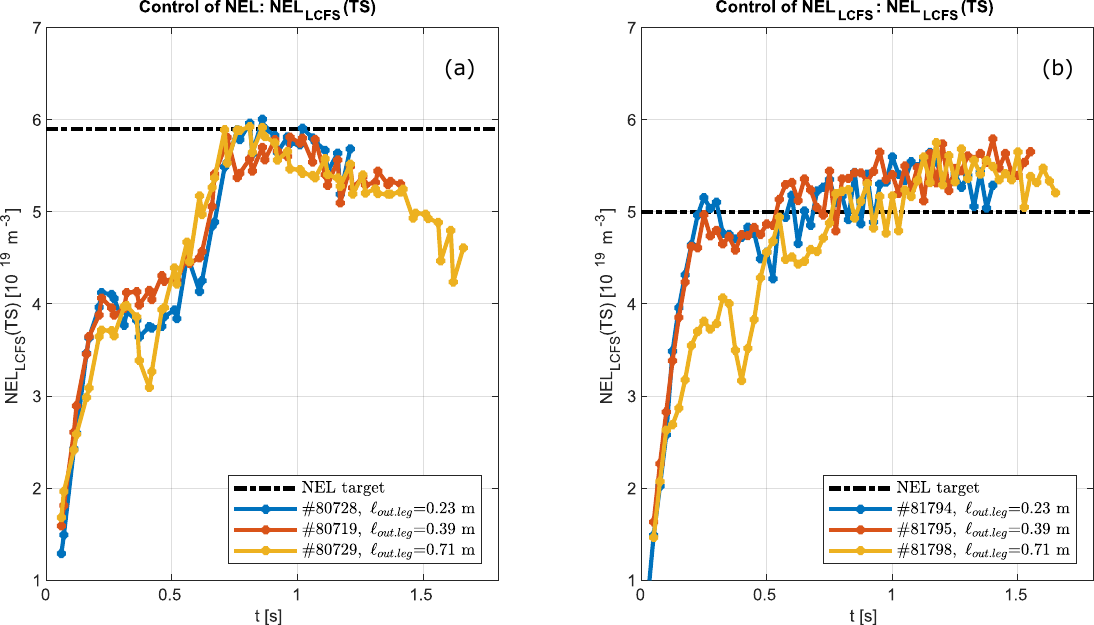}
  \centering
  \caption{$\mathrm{NEL}_{\mathrm{LCFS}}$ computed with offline TS data and offline LIUQE geometry computation. On the left, density control performed with the traditional feedback loop, on the right, density control performed with the RAPDENS-based loop. The color code follows the one adopted in Figure \ref{fig:liuqe_TS_FIR}. Target density in terms of $\mathrm{NEL}_{\mathrm{LCFS}}$ is shown in dashed black.}
  \centering
  \label{fig:RAPDENS-based control of NEL,LCFS}
\end{figure}
\clearpage

\begin{figure}[h!]
\centering
  \includegraphics[width=0.95\textwidth]{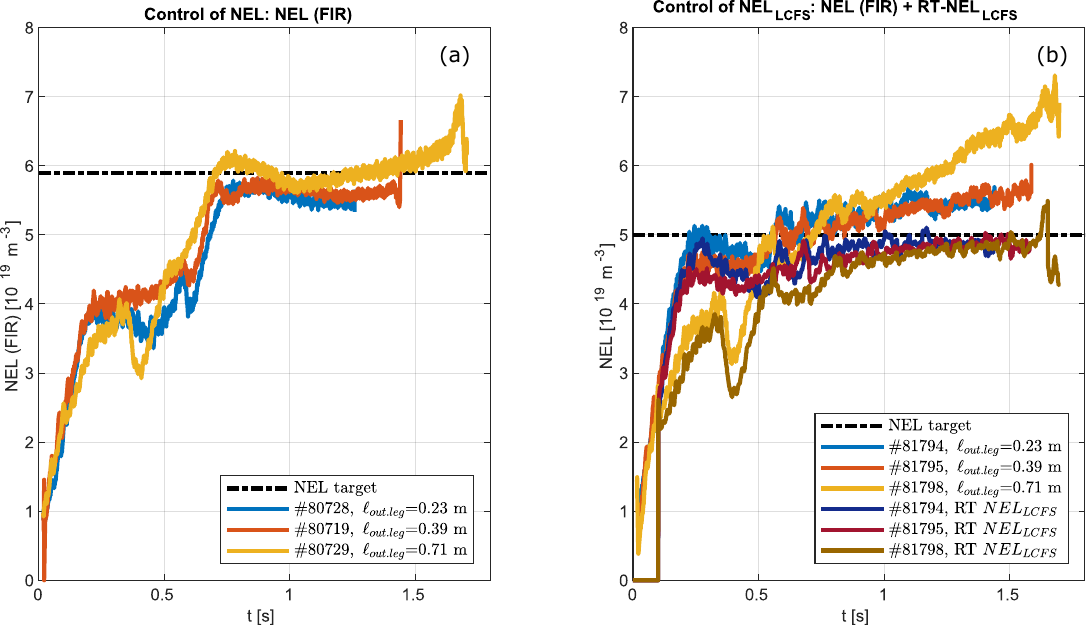}
  \centering
  \caption{Comparison between the line-averaged electron density $\mathrm{NEL}_{\mathrm{FIR}}$ computed with the central FIR channel in the set of shots scanning the outer divertor leg length: on the left, density control performed with the traditional feedback loop, on the right, density control performed with the RAPDENS-based loop. The color code follows the one adopted in Figure \ref{fig:liuqe_TS_FIR} to indicate the various divertor configurations. The dashed line in black shows the target $\mathrm{NEL}_{\mathrm{LCFS}}$, as a reference. The real-time signals employed for the control of $\mathrm{NEL}_{\mathrm{LCFS}}$ based on RAPDENS reconstruction are reported in darker shades for the three divertor configurations.}
  \centering
  \label{fig:RAPDENS-based control of NEL,FIR}
\end{figure}

\begin{figure}[h!]
\centering
  \includegraphics[width=0.95\textwidth]{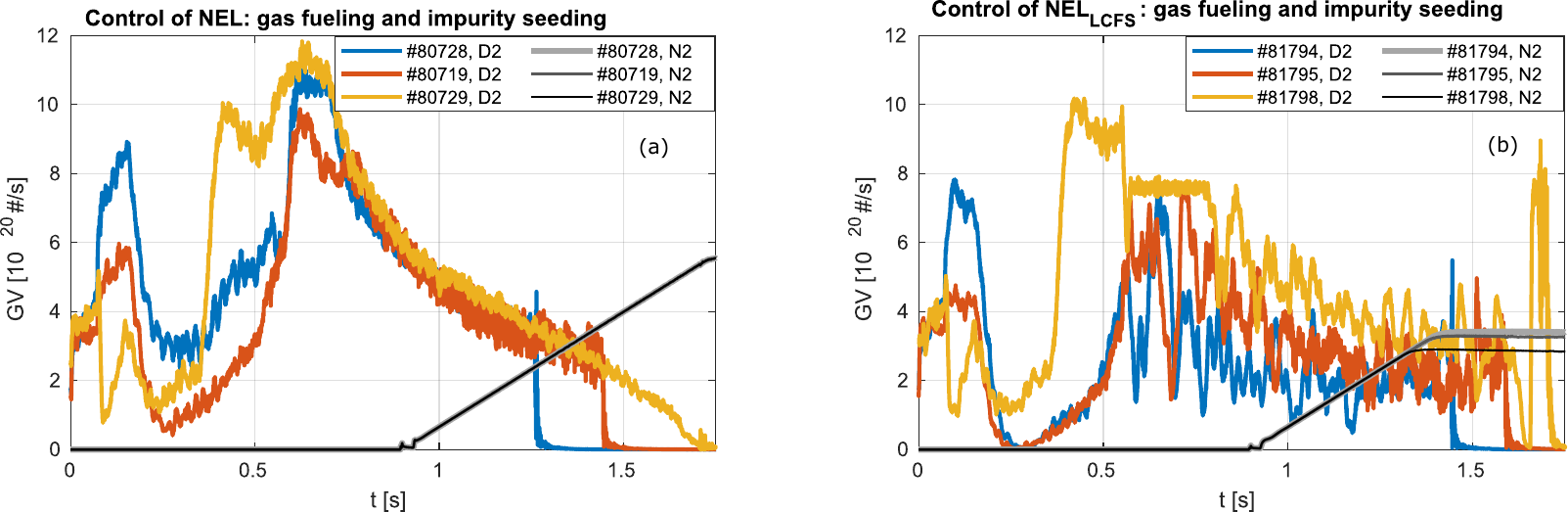}
  \centering
  \caption{Gas valve fueling and impurity seeding for traditional density control of NEL, Figure \ref{fig:gas_leg_scan}a, and RAPDENS-based density control of $\mathrm{NEL_{LCFS}}$, Figure \ref{fig:gas_leg_scan}b. Color code for $D_2$ gas traces follows the one used in Figure \ref{fig:liuqe_TS_FIR}. The feedforward traces for $N_2$ impurity seeding are represented in scales of gray. A saturation of the impurity seeding flow occurs at different levels in \ref{fig:gas_leg_scan}b, due to insufficient upstream gas pressure.}
  \centering
  \label{fig:gas_leg_scan}
\end{figure}
\section{Control of the central density in ECH and NBH discharges}\label{section: ne control below cutoff}
Stable high-performance tokamak operation requires the simultaneous fulfillment of multiple tasks, including efficient plasma heating and current drive. To this end, a combination of heating schemes based on energetic neutral particles and microwave power is employed, notably ECH and ECCD, together with Ion Cyclotron Heating (ICH) and NBH, as foreseen for ITER and DEMO \cite{Jelonnek2017}. ECH and ECCD offer highly localized power deposition and non-inductive current drive. In particular, ECCD enables extended pulse operation with reduced magnetic flux consumption and can be applied off-axis to tailor the safety factor profile for advanced scenarios \cite{sauter_steady-state_2000,Coda_2000,piron_extension_2019} and to suppress MHD instabilities such as neoclassical tearing modes \cite{Kong_2022}. To ensure effective coupling between plasma and EC microwaves, it is required to keep the electron density below the cutoff limit, which is fixed by the gyrotron frequency, harmonic, polarization, and magnetic field intensity in the tokamak. For TCV, in the case of EC X2 heating, the cutoff density threshold is $\mathrm{n_{e} = 4.0 \cdot 10^{19} \ m^{-3}}$. For this purpose, the RAPDENS observer can be leveraged to reconstruct the local features of the electron density profile and propagate the estimate of the density at a radial coordinate $\mathrm{\rho=\rho_{target}}$ to be kept below the cutoff limit, ensuring a full absorption of the EC wave on the resonance spatial location. 


The experimental objective is to locally control a point of the electron density profile on the $\rho$ grid, while preventing ECH cutoff, by providing a reference density trace reconstructed by RAPDENS to a PI controller within SAMONE to modulate the injection of fueling gas, as different heating mixes act as disturbances to the task. The target density is chosen to lie below the cutoff at the spatial point where the ECH resonant location is, and the tracking PI controller is used to reach the target and react in case of disturbances from heating sources or the wall recycling action.  


\subsection{Control of $n_e(\rho=0)$ during on-axis X2-EC heating, shot \#82913 }\label{section:control_ne0_82913}
A first experiment to test the new control scheme in the presence of on-axis ECH power has been carried out in shot \#82913. ECH in X2 is applied at the center of the plasma, and the central electron density, reconstructed by RAPDENS, is feedback-controlled to follow a reference below the density cutoff. The main time traces are reported in Figure \ref{fig:ne core control 82913}.

\begin{figure}[h!]
\centering
  \includegraphics[width=0.95\textwidth]{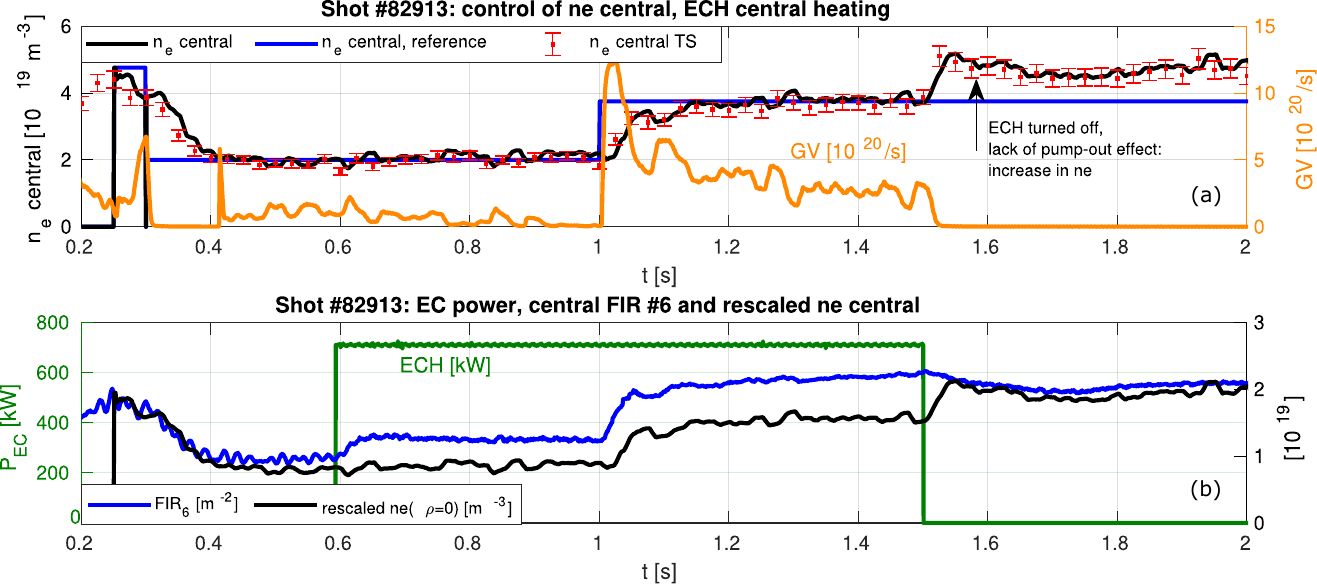}
  \centering
  \caption{Control of $n_e(\rho=0)$ with gas valve feedback control, shot \#82913. On the top figure, the reconstructed $n_e(\rho=0)$ signal with RAPDENS observer is shown in black, the reference target trace in blue, offline TS points in red, and the injected gas flux in orange. On the bottom, ECH power, in green, the FIR signal from channel \#6, in blue, and the time trace of $n_{e}(\rho=0)$, rescaled over the value of FIR \#6 at $t=0.250 \ s$, in black.}
  \centering
  \label{fig:ne core control 82913}
\end{figure}

In Figure \ref{fig:ne core control 82913}a, the signals related to the control task are summarized. RAPDENS reconstructed central electron density $\mathrm{n_{e,central}=n_{e}(\rho=0)}$ (in black), TS data points (in red), target density (in blue), and gas valve time traces (in yellow) are reported. In Figure \ref{fig:ne core control 82913}b, the injected ECH power is shown (in green), together with the time traces of the FIR $\#6$ line integral $\mathrm{y_{\mathrm{FIR},6}}$ (in blue) and the normalized timetrace $\mathrm{n_{e}(\rho=0)/y_{\mathrm{FIR},6}(t=0.250 \ s)}$ (in black). The latter quantity is quite insightful in capturing changes in the density peaking throughout the experiment.

There is good agreement, within error bars, between the reconstructed central density and the offline TS value, indicating that the observer can locally track the density profile by means of the RAPDENS predictive model, FIR, and TS measurements. SAMONE density feedback control is activated at $\mathrm{t = 0.250 \ s}$ with the reference set to the density value at the time of controller switch-on. Then, at $\mathrm{t = 0.300 \ s}$, the reference is changed to the absolute value of $\mathrm{n_{e,target} (\rho=0)= 2.0 \cdot 10^{19} \ m^{-3}}$. In this phase, the controller instructs the complete closure of the gas valve, and the central density decreases from $4.0$ to $\mathrm{2.0 \cdot 10^{19} \ m^{-3}}$, in $\mathrm{\approx 100 \ ms}$. The central density remains well controlled during the time interval $\mathrm{t=[0.400 \ s , 1.00 \ s]}$. In this phase, the ECH is injecting 700 kW into the core of the plasma. The delivered $D_2$ gas flux to sustain this section of the control experiment is around $ \mathrm{0.700 \cdot 10^{20} \ molec./s}$.

The impact of the injection of ECH can be noticed on the bottom plot of Figure \ref{fig:ne core control 82913}, where the ratio of the central FIR \#6 signal over the rescaled $\mathrm{n_e(\rho=0)}$ increases. This variation in the ratio suggests that a flattening of the density profile and/or an increase in $\mathrm{NEL}_{\mathrm{SOL}}$ occurs in this ECH phase, since $\mathrm{n_e(\rho=0)}$ is kept constant while the integral value measured by the FIR \#6 on the subtended area of the $\mathrm{n_e}$ profile increases. More details on this are reported in Section \ref{section:ne_profile_heating_schemes}. A step increase of the target density, from $2.0$ to $\mathrm{3.7 \cdot 10^{19} \ m^{-3}}$, has been applied at t = 1.00 s. The controller reacts by commanding more fueling gas to let the controlled $\mathrm{n_{e,central}}$ reach the new target, which is approached in 100 ms and further sustained in the time window $\mathrm{t=[1.10 \ s, 1.50 \ s]}$. The gas flux injected stabilizes at around $ \mathrm{2.4 \cdot 10^{20} \ molec./s}$. Full-power absorption of the EC wave has been verified with post-shot TORAY-GA \cite{matsuda_ray_1989} standard analysis throughout the ECH section of the shot. The ECH power is then turned off at t=1.50 s, and an abrupt increase in central density is verified. The controller completely closes the gas valve to counteract the sudden increase in density. Even in the absence of external fueling, the central density does not decrease towards the target. Notably, as shown in Figure \ref{fig:ne core control 82913}b, the line integral value of the density, measured with FIR \#6, slightly decreases and stabilizes at a value of $\mathrm{2.0 \cdot 10^{19} \ m^{-2}}$, suggesting that a change in the peaking factor occurred, recovering the ohmic profile before the EC injection albeit at a higher density, while the total particle amount is maintained constant in this final section of the shot through wall recycling. 

The real-time estimation of the electron pinch velocity-to-diffusivity ratio for this specific scenario, as described in Section \ref{section:estimation_nu_TS}, is reported in Figure \ref{fig:nu 82913 timetraces}. The different phases of the shot, in limited, diverted ohmic (before and after ECH), and diverted ECH, are characterized by three distinct profiles, captured by the electron pinch velocity estimation algorithm. The profiles are reported in Figure \ref{fig:nu 82913 timeslices}. It is interesting to note that the profile shape, captured by $\nu_{\mathrm{TS}/D}$, is modified with a delay of $\mathrm{\tau_{n}\approx100 \ ms}$ dictated by the particle confinement time for each phase, as shown in Figure \ref{fig:nu 82913 timetraces}. In addition, the ECH phase of the shot is characterized by lower $\nu$ values and an outward electron pinch velocity (blue trace in Figure \ref{fig:nu 82913 timetraces}, corresponding to $\mathrm{\nu_{\mathrm{TS}}/D(\rho=0.20)}$) leading to a flattening of the profile.

\begin{figure}[h!]
\centering
  \includegraphics[width=0.75\textwidth]{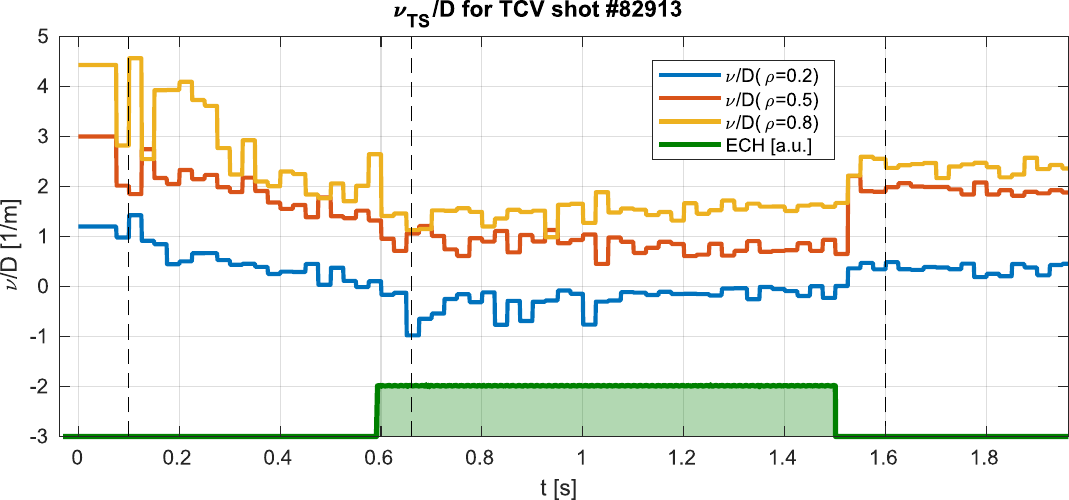}
  \centering
  \caption{ Real-time estimation of the electron pinch velocity-to-particle diffusivity ratio $\nu/D$ with the EKF algorithm in shot \#82913, adopting the L-mode diffusion coefficient shown in Figure \ref{fig:diffusion_coeff}. Different timetraces are reported at the radial coordinates $\rho\in[0.20,0.50,0.80]$. Limited phase in $\mathrm{t\in[0 \ s,0.260\ s]}$. The ECH phase of the shot is highlighted with a green box. Dashed black lines indicate the timeslices location reported in Figure \ref{fig:nu 82913 timeslices}.}
  \centering
  \label{fig:nu 82913 timetraces}
\end{figure}

\begin{figure}[h!]
\centering
  \includegraphics[width=0.55\textwidth]{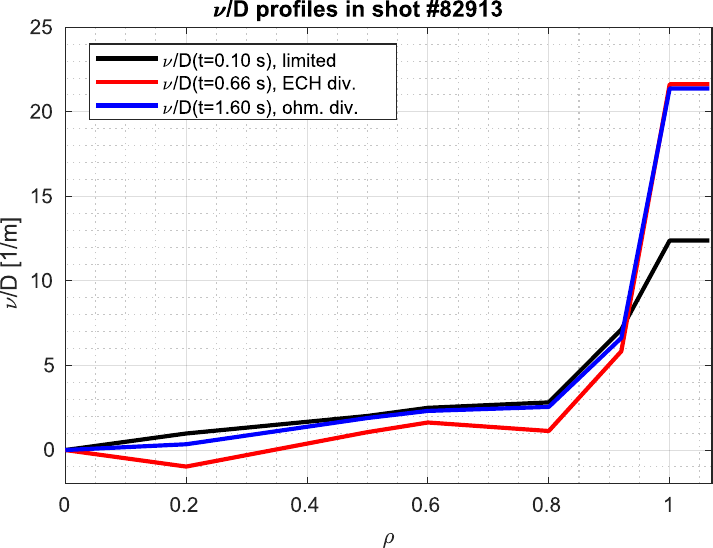}
  \centering
  \caption{  Real-time estimation of the electron pinch velocity-to-particle diffusivity ratio $\nu/D$ in shot \#82913. Ohmic, limited profile in black, diverted, ECH profile in red, and ohmic, diverted profile in blue. The profiles are taken in correspondence with the vertical dashed lines of Figure \ref{fig:nu 82913 timetraces}.}
  \centering
  \label{fig:nu 82913 timeslices}
\end{figure}

A hypothesis on the particle behavior observed when the ECH is turned off at t=1.50 s can be proposed based on the density pumpout effect \cite{Weisen2001} driven by ECH injected power, where core particles are expelled with an enhanced transport driven by Trapped Electron Mode (TEM) instabilities.
The sudden switching off of the ECH power lowers the TEM contribution, yielding stronger ITG and inward pinch, which is reflected in an abrupt change in the $n_e$ peaking profile (and the opposite is observed when ECH is turned on). The plasma density fueling provided by wall recycling maintains the central and averaged density constant over the last section of the shot.
To further expand on this, a turbulent simulation using the GENE code \cite{GENE} on different instants of the shot, in the ohmic phase (t= 0.500 s), ECH phase (t=1.40 s), and the ohmic phase after the ECH (t=1.80 s) has been conducted. The details on the simulation setting and main results are presented in the  Section \ref{section:GENE_Results}.

\subsection{Control of $n_e(\rho=0)$ in the heating mix of on-axis X2-EC, NBI-1 and NBI-2, shot \#82893}
A further step to test the control scheme has been performed in shot $\#82893$, where the influence of NBI heating in co-current (NBI-1) and counter-current (NBI-2) on the density control task is shown in Figure \ref{fig:ne core control 82893}.

\begin{figure}[h!]
\centering
  \includegraphics[width=\textwidth]{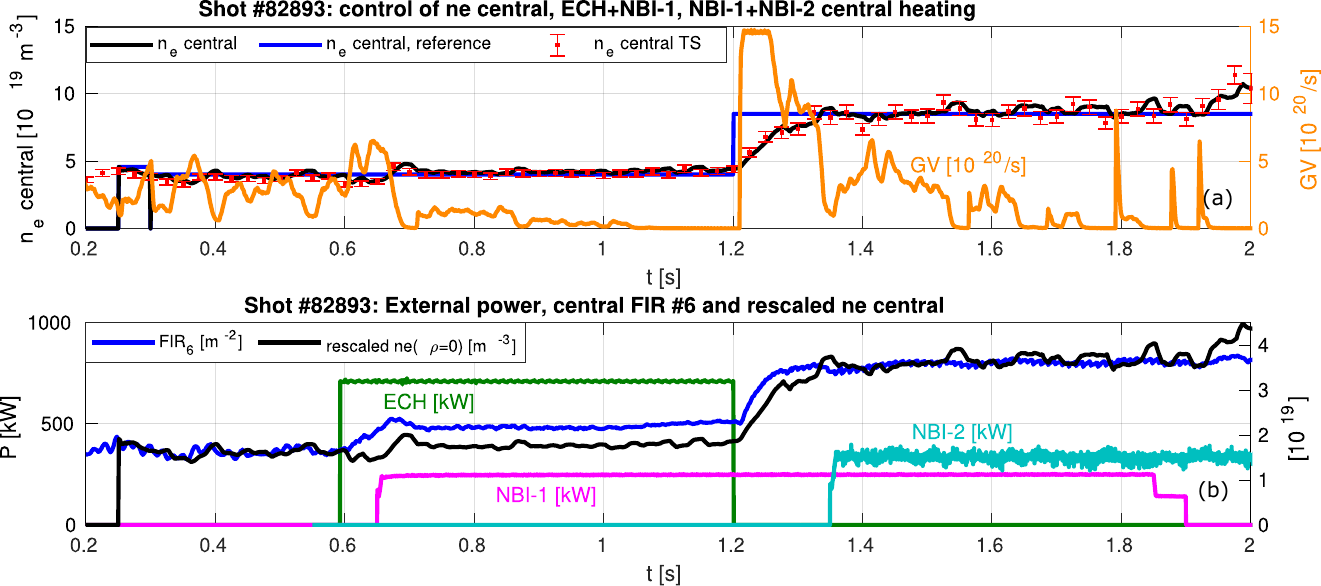}
  \centering
  \caption{Control of $n_e(\rho=0)$ with gas valve feedback control, shot \#82893. On the top figure, the reconstructed $n_e(\rho=0)$ signal with RAPDENS observer is shown in black, the reference target trace in blue, offline TS points in red, and gas valve command in orange. On the bottom, the applied ECH power in green, the NBI-1 power in magenta, and the NBI-2 power in cyan are represented. The FIR signal from channel \#6 is shown in blue, and the rescaled value of $n_{e}(\rho=0)$ over the value of FIR \#6 at $\mathrm{t=0.250 \ s}$ in black.}
  \centering
  \label{fig:ne core control 82893}
\end{figure}

The experimental setup is based on shot \#82913, where the same equilibrium, X2-ECH launching angle and power, and controller gains are adopted. The target density is now set at $\mathrm{4.0 \cdot 10^{19} m^{-3}}$ for $\mathrm{t=[0.300 \ s ,1.20 \ s ]}$ to demonstrate robust control close to cutoff conditions. 
NBI-1 heating (in purple, Figure \ref{fig:ne core control 82893}b) is applied in $\mathrm{t = [0.650 \ s, 1.90 \ s]}$, with an injection of 300 kW of power. 
In this first section of the shot, as shown in Figure \ref{fig:ne core control 82893}a, the controlled $\mathrm{n_{e,central}}$ is sustained close to the target. A small excursion can be noticed around $\mathrm{t= 0.600 \ s}$ when ECH power is injected, leading to an undershoot of the density. At $\mathrm{t\approx0.700 \ s}$, an overshoot, quickly under controlled, is instead driven by the injection of NBI-1, where the ionization of the injected particles acts as an extra source of particles in the plasma core. The controller reacts to the injection of NBI-1 by rapidly decreasing the flux from $7.00$ to $\mathrm{0.700 \cdot 10^{20} \ molec./s}$. From this point on, the $D_2$ flux required to sustain this plasma density level decreases over time, from $\mathrm{1.27 \cdot 10^{20} \ molec./s}$ to complete closure of the gas valve, due to the increasing effect of neutral recycling from the wall. 

The ECH phase of the shot ends at $\mathrm{t=1.20 \ s}$, and a step-wise modification in the target density is introduced, setting up a higher value of $\mathrm{8.5 \cdot 10^{19} \ m^{-3}}$. The updated target has been introduced to enable the controller to sustain a density level requiring a non-zero gas puffing, to avoid the behavior shown in the last ohmic section ($\mathrm{t\geq1.50 \ s}$) of shot $\#82913$, in Figure \ref{fig:ne core control 82913}a. The value $\mathrm{8.5 \cdot 10^{19} \ m^{-3}}$ has been chosen based on previous development shots, $\#82889$ and $\#82892$, not shown here for brevity. The central density reaches the new target in $\mathrm{\approx 150 \ ms}$, and the PI controller gas flux keeps the central density sustained for the rest of the control task. NBI-2 is fired at t=1.40 s, where 350 kW of power is injected in the counter-current direction (in cyan). A final section of the shot, at t=1.90 s, features only NBI-2 heating, and the density exhibits an increase due to the absence of sawteeth activity in the core (examined via the soft-X ray diagnostic), indicating a change of the global q profile leading to disruption. Figure \ref{fig:ne core control 82893}a shows a good agreement of the reconstructed on-axis $\mathrm{n_{e}}$ with offline TS measurements, within error bars.

In Figure \ref{fig:ne core control 82893}b, the ratio between the FIR \#6 and $n_e(\rho=0)$ is represented. Similarly to the result presented in Section \ref{section:control_ne0_82913}, an increase in the ratio between the central interferometer and the reconstructed $n_e(\rho=0)$ is evident when ECH and NBI-1 are inserted, at $t=0.650$ s. The ratio changes in the NBH case, stabilizing to values comparable with the ohmic phase at $\mathrm{t=[0.200 \ s,0.600 \ s]}$.

\subsection{Effect of the heating schemes on the density profile}\label{section:ne_profile_heating_schemes}
To further investigate the performance of RAPDENS profile reconstruction in this diverse mix of heating schemes and scenarios, a series of time-slices of the density profile in different sections of the shots $\#82913$ and $\#82893$ is reported in Figure \ref{fig:ne profiles TS 82893 82913}, compared with offline TS data. The estimation of the electron pinch velocity-to-diffusivity ratio $\nu_{\mathrm{TS}/D}$ is also presented.

\begin{figure}[h!]
\centering
\hspace{-1.5cm}
  \includegraphics[width=0.95\textwidth]{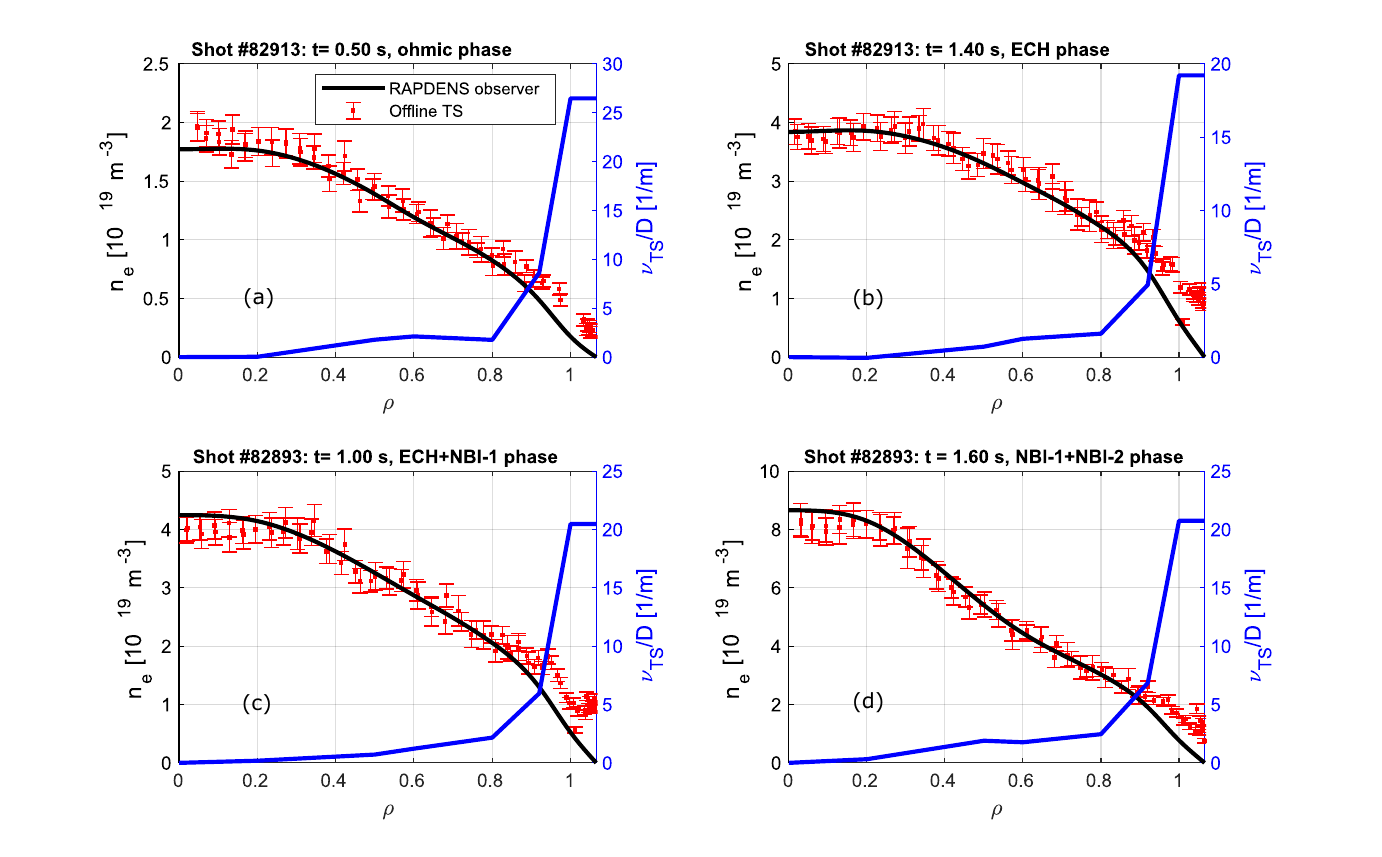}
  \centering
  \caption{Snapshots of RAPDENS reconstruction of the $n_e$ profile, in black, and offline TS data in red, for shots \#82913 and \#82893. The estimated electron pinch velocity-to-diffusivity coefficient $\nu_{\mathrm{TS}}/D$ is reported in blue, assuming a fixed particle diffusion coefficient $D=D(\rho)$, shown in Figure \ref{fig:diffusion_coeff}. A good agreement of the reconstructed profile over TS points is found, up to the coordinate $\rho\approx0.80$. A systematic underestimation of the profile at the edge can be observed mainly due to the adopted boundary condition, see Equation (\ref{eq:BC_ne}).}
  \centering
  \label{fig:ne profiles TS 82893 82913}
\end{figure}

The reconstructed profiles lie within the TS error bars in the four reported snapshots, up to a value of $\rho\approx0.80$. However, an underestimation of the density at the edge, for values of $\mathrm{\rho>0.80}$, is evident in all four cases. This reconstruction deficiency primarily arises from the enforcement of the homogeneous Dirichlet boundary condition $\mathrm{n_{e}(\rho=\rho_e)=0}$, as shown in Equation (\ref{eq:BC_ne}). It is, in fact, difficult to predict the value of $\rho_e$ at which it would allow matching the profile near $\mathrm{\rho=1.0}$. In addition, $\mathrm{n_{e}}$ is never null in the SOL. To improve such reconstruction behavior, a new set of boundary conditions developed in \cite{Kropackova2025} has been tested using an offline simulation with real-time data inputs and measurements, to enhance the accuracy of the observer in the edge region. However, these were not available at the time of these experiments. The separatrix density $n_e(\rho=1.0)$ can be estimated from empirical formulas \cite{Kallenbach2018}, SOL reduced models, or direct estimation through RT-TS data, and enforced directly with the novel numerical formulation of the boundary conditions present in the RAPDENS predictive model. In this work, the latter has been considered in the approach due to the availability of RT-TS measurements in the control system. The offline comparison between the reconstruction performances of the two different sets of boundary conditions is reported in \ref{section:Appendix_inhomog_BC}, with a clear improvement of the reconstruction in the edge region $\mathrm{\rho\in[0.80,1.0]}$.

To further analyze the impact of the different heating schemes on the density profile, a set of averaged profiles, obtained from the RAPDENS reconstruction in various sections of the shots $\#82893$ and $\#82913$ has been considered. The profiles have been normalized to the central electron density, as shown in Figure \ref{fig:ne profiles peaking 82893 82913}. The ohmic profile, represented in black, is used as a reference for the three different cases. The shaded area represents the standard deviation derived from the set of density profiles considered in the averaging procedure, computed for each radial coordinate. For the ohmic case, the averaging time window is chosen as $\mathrm{t_{ohm}=[0.420 \ s,0.600 \ s]}$, based on shot \#82913. The selected time window corresponds to the shot section where the central density is controlled close to the density target. In Figure \ref{fig:ne profiles peaking 82893 82913}a, the ohmic case is compared with the pure ECH case, averaging over $\mathrm{t_{ECH}=[0.750 \ s, 1.00 \ s]}$ of shot \#82913. Recalling what was anticipated in the Section \ref{section:control_ne0_82913}, an important change in peaking factor can be isolated. The flattened density profile at $\rho=0$ and the increased subtended area are reflected in a higher value of the FIR \#6 signal, compared with the ohmic case, yielding a lower density scale length mid-radius. The flattened density profile derives from a more outward electron pinch transport term, corresponding to the particle density pumpout effect. Vice versa, in Figure \ref{fig:ne profiles peaking 82893 82913}b, a dual behavior in the density profile peaking is present. A highly peaked profile is evident in comparison with the ohmic one. This is in part due to the localized ionization of particles injected inside the plasma core with the NBI-1 and -2 and to the increase of the ITG inward pinch contribution favoured by NBI (ion) heating. The high peaking of the profile translates to a lower value of the subtended area than the ohmic case, which, in turn, reflects in a reduction of the FIR \#6 to $n_{e}(\rho=0)$ ratio, as reported in Figure \ref{fig:ne core control 82893}b, in the time interval $\mathrm{t_{NBI-1+NBI-2}=[1.40 \ s,1.90 \ s]}$. The similarity of the FIR-to-$n_{e}(\rho=0)$ ratio to the value computed in the ohmic section of the shot can be motivated by two effects acting in opposite directions: the lower subtended area, on one hand, compensated, on the other, with an increased pickup of the SOL density in the divertor. In Figure \ref{fig:ne profiles peaking 82893 82913}c, an interesting effect on the mixed heating case (ECH and NBI-1) can be highlighted. The profile, averaged over the time window $\mathrm{t_{ECH+NBI-1}=[0.800 \ s,1.10 \ s ]}$, retains the spatial features of the ohmic case. The effect of the ECH and NBI-1 actuators on the profile compensates one another, promoting outward pinch with ECH and inward pinch with NBI.

These analyses and studies can establish a basis for designing a profile controller, where the different actuators can be leveraged to adjust the peaking factor of the profile in real-time. The effects of the actuators on the transport coefficients or source profile term can be incorporated into the RAPDENS predictive model and utilized for the design of an MPC, as done in \cite{BosmanITER}. These experiments allow to test the effect of heating mix on particle transport at same central density, which is discussed in the next Section.

\begin{figure}[h!]
\centering
\hspace{-1cm}
  \includegraphics[width=0.95\textwidth]{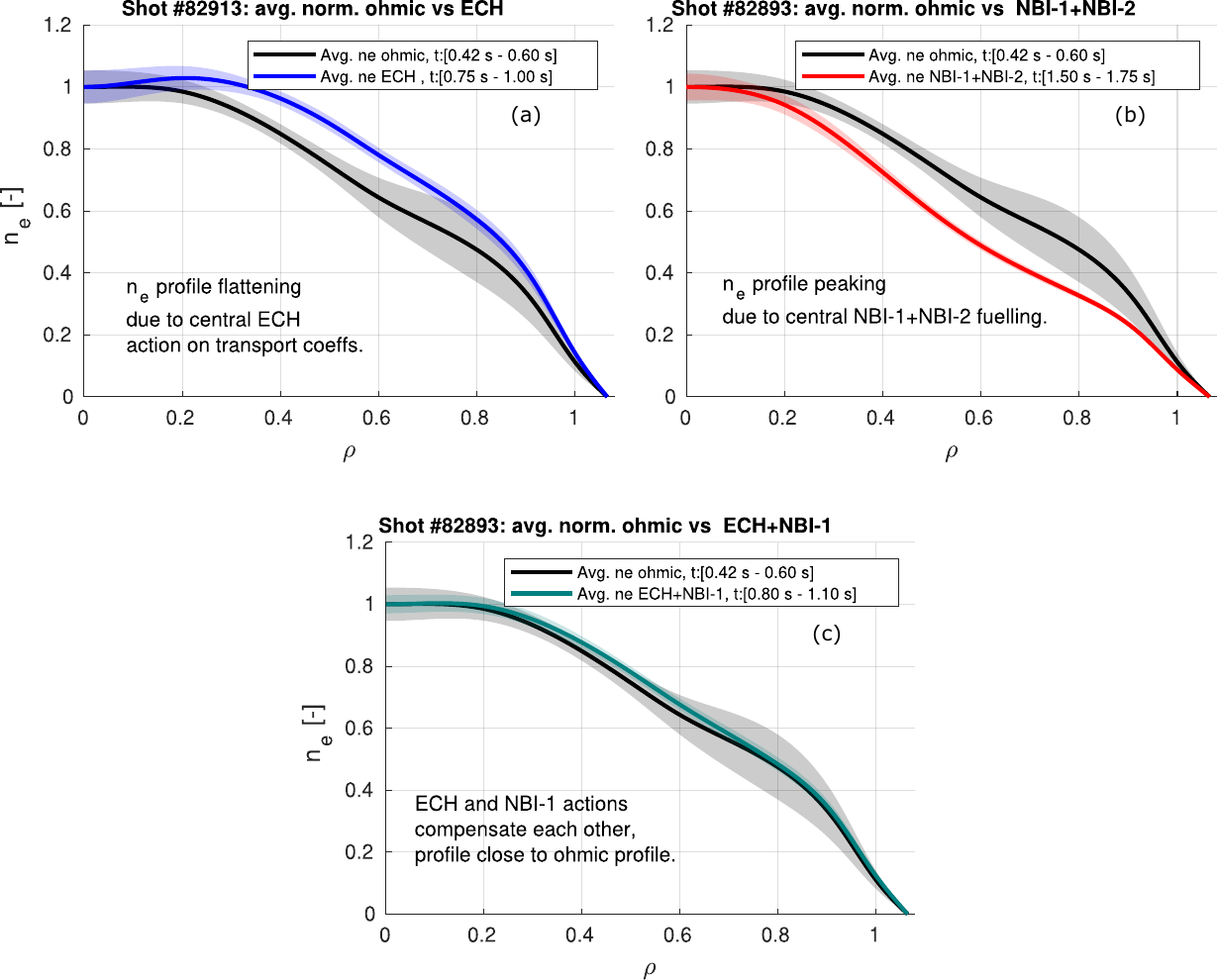}
  \centering
  \caption{Snapshots of RAPDENS reconstruction of $n_e$ over different sections of the shots \#82913 and \#82893. The profiles are normalized over $n_e(\rho=0)$ and averaged over time windows related to stationary sections of the shot. The shaded area represents the standard deviation of the reconstructed profiles for each radial coordinate. Different peaking factors on $n_e$ can be isolated, compared to the reference ohmic profile reported in black.}
  \centering
  \label{fig:ne profiles peaking 82893 82913}
\end{figure}

\subsection{Impact of the adaptive $\nu/D$ ratio in the spatial reconstruction for shot \#82913}\label{section:impact_nu_D_82913}
To quantify the impact of the adoption of the time-varying $\nu_{TS}/D$ ratio in the estimation of the electron density profile, a comparison of the estimation of the profile with and without such time-adapting technique is presented. An offline simulation with fixed $\nu$ value tuned heuristically, using the same RT data and settings for the covariance matrices, is carried out and compared with the results shown in the previous Sections. The simulation employs the same particle diffusion coefficient. The result is shown in Figure \ref{fig:ne nu adaptive 82913}, where the reconstructed density profiles, in blue and black, representing the estimation respectively with and without the adaptive $\nu_{TS}/D$, are compared with TS data, in red. Overall, a non-negligible impact of the estimation of $\nu_{TS}/D$ in the region $\mathrm{\rho\in[0,0.20]}$ can be appreciated, due to the capability of the proposed estimation algorithm to capture possible variations in the peaking factor and changes in monotonicity of the core density profile, due to sawteeth instabilities or strong outwards pinch transport. The profiles estimated with such a time-adaptive technique result closer to the TS experimental data and thus provide more accurate results in such experimental conditions.

\begin{figure}[h!]
\centering
\hspace{-1cm}
  \includegraphics[width=0.75\textwidth]{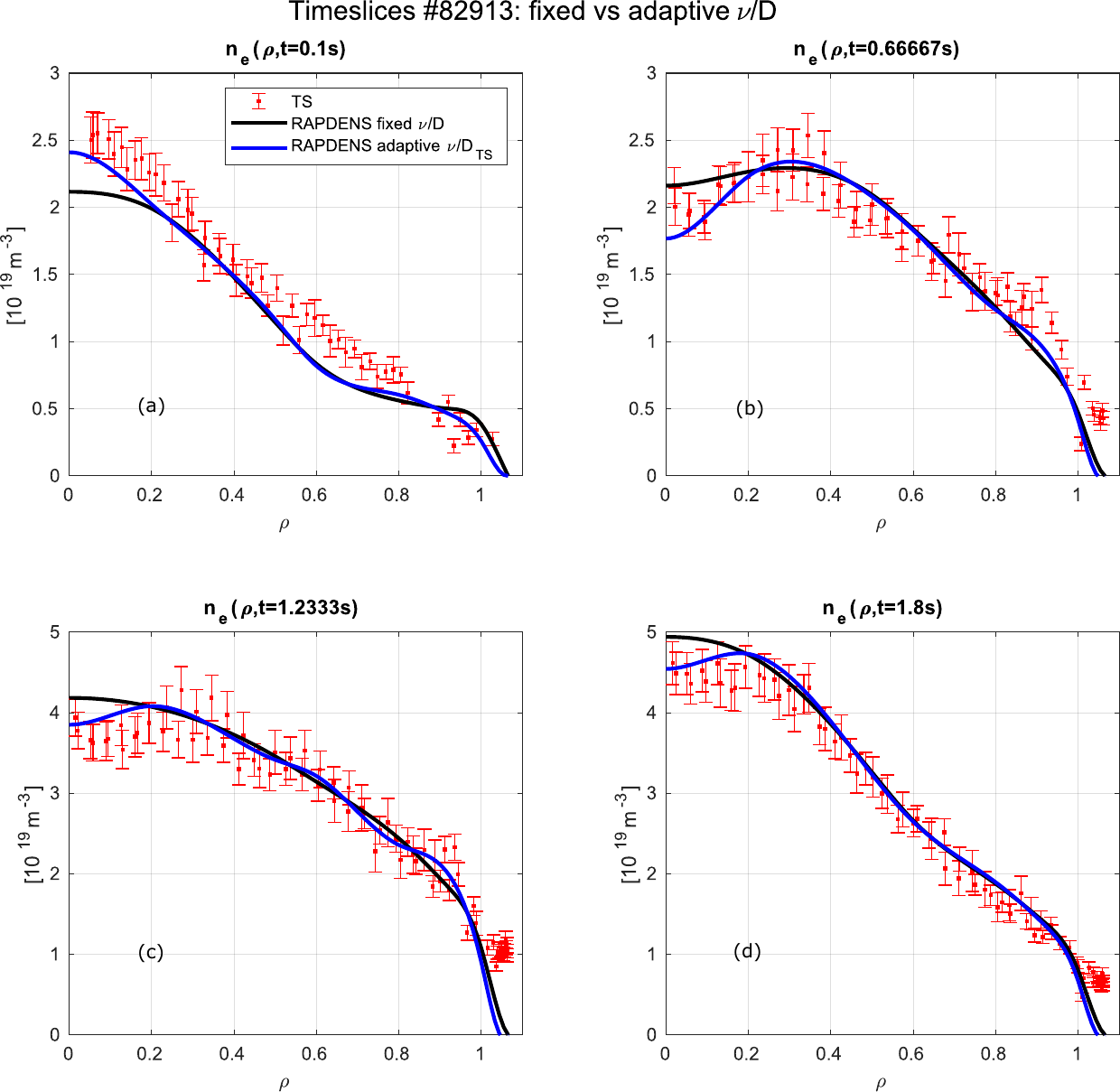}
  \centering
  \caption{Comparison of the fixed vs time-adaptive timeslices for RAPDENS reconstruction of $n_e$ over different sections of the shot \#82913. Fixed ratio $\nu/D$ in black, time-adaptive estimation ratio in blue. TS data in red, with error bars. }
  \centering
  \label{fig:ne nu adaptive 82913}
\end{figure}

\subsection{Turbulent transport characterization of shot \#82913 with GENE simulations}\label{section:GENE_Results}
To assess how turbulent transport influences the required gas to maintain the reference central density in shot \#82913, gyrokinetic simulations have been performed with the flux tube gradient-driven version of the GENE code \cite{GENE}. GENE is a physically comprehensive Eulerian gyrokinetic code that solves the 5-dimensional discretized Vlasov-Maxwell equations. The code can be run in a linear or non-linear mode. The former is much less computationally expensive and is especially useful to develop an understanding of the nature of the turbulence that dominates the system. The latter is considerably more expensive, but is needed to have a physically accurate model of micro-turbulence. Indeed, one must consider the non-linear interaction for turbulence to saturate and to assess the transport level. To perform the simulations, a tracer-efit algorithm \cite{xanthopoulos_geometry_2009} was used to reconstruct the magnetic equilibrium at the desired radial location (i.e. $\mathrm{\rho=0.75}$) from EQDSK files of the experiments. 

The simulations were performed at $\rho=0.75$ for the three considered times (t=0.500 s, t=1.40 s and t=1.80 s) of shot \#82913. Table \ref{input} shows the key parameters used to run the simulations.

\begin{table}[h]
    \centering
    \begin{tabular}{l|ccc}
    \toprule
        &$t=0.500 \ s$ & $t=1.40 \ s$ & $t=1.80 \ s$ \\ \hline
        $\rho$& 0.75 & 0.75 & 0.75\\
        $a/L_{Te}$& 3.75 & 4.05 & 4.23\\ 
        $a/L_{Ti}$& 1.74 & 1.21 & 1.45\\ 
        $a/L_{ne}$& 1.82 & 1.80 & 2.18\\
        $a/L_{nC}$& 2.98 & 2.81 & 1.65 \\
        $T_e$ [keV]& 0.23 & 0.28 & 0.18\\ 
        $T_i/T_e$& 0.66 & 0.73 & 1.15 \\ 
        $n_e\,[10^{19}m^{-3}]$& 0.99 & 2.52 & 1.96 \\ 
        $n_C/n_e$& 0.04 & 0.03 & 0.04 \\\toprule
    \end{tabular}
    \caption{\label{input} Key simulation parameters for various experimental scenarios including the logarithmic gradients of electron temperature $R/L_{Te}$, ion temperature $R/L_{Ti}$, electron density $R/L_{ne}$, carbon density $R/L_{nC}$, as well as the ion-electron $T_i/T_e$ temperature ratio, electron density $n_e$, the carbon to electron density $n_C/n_e$. Inverse scale lengths for temperature and density profiles computed as $R/L_{T} = \dfrac{\partial}{\partial\rho}log({T})$. Electron kinetic information extracted from TS data, main ions, and carbon kinetic profiles from the CXRS system \cite{bagnato_study_2024}. }
\end{table}

Linear simulations, not shown here for brevity, show a TEM-dominated regime for all three scenarios. non-linear simulations have been performed to study how particle flux changes in the three scenarios. Figures \ref{Q} and \ref{G} show, respectively, the heat and particle fluxes for the three scenarios computed with non-linear simulations. The heat fluxes show that the largest component is the electron electrostatic heat flux, confirming that TEM is dominating transport. Moreover, a comparison with the heat fluxes computed with ASTRA \cite{ASTRA} interpretative simulations shows good agreement, thus suggesting that these simulations obtained a good level of reliability, as shown in Figure \ref{Q} (error bars represent the experimental values).

\begin{figure}[h!]
    \centering
    \includegraphics[width=0.75\linewidth]{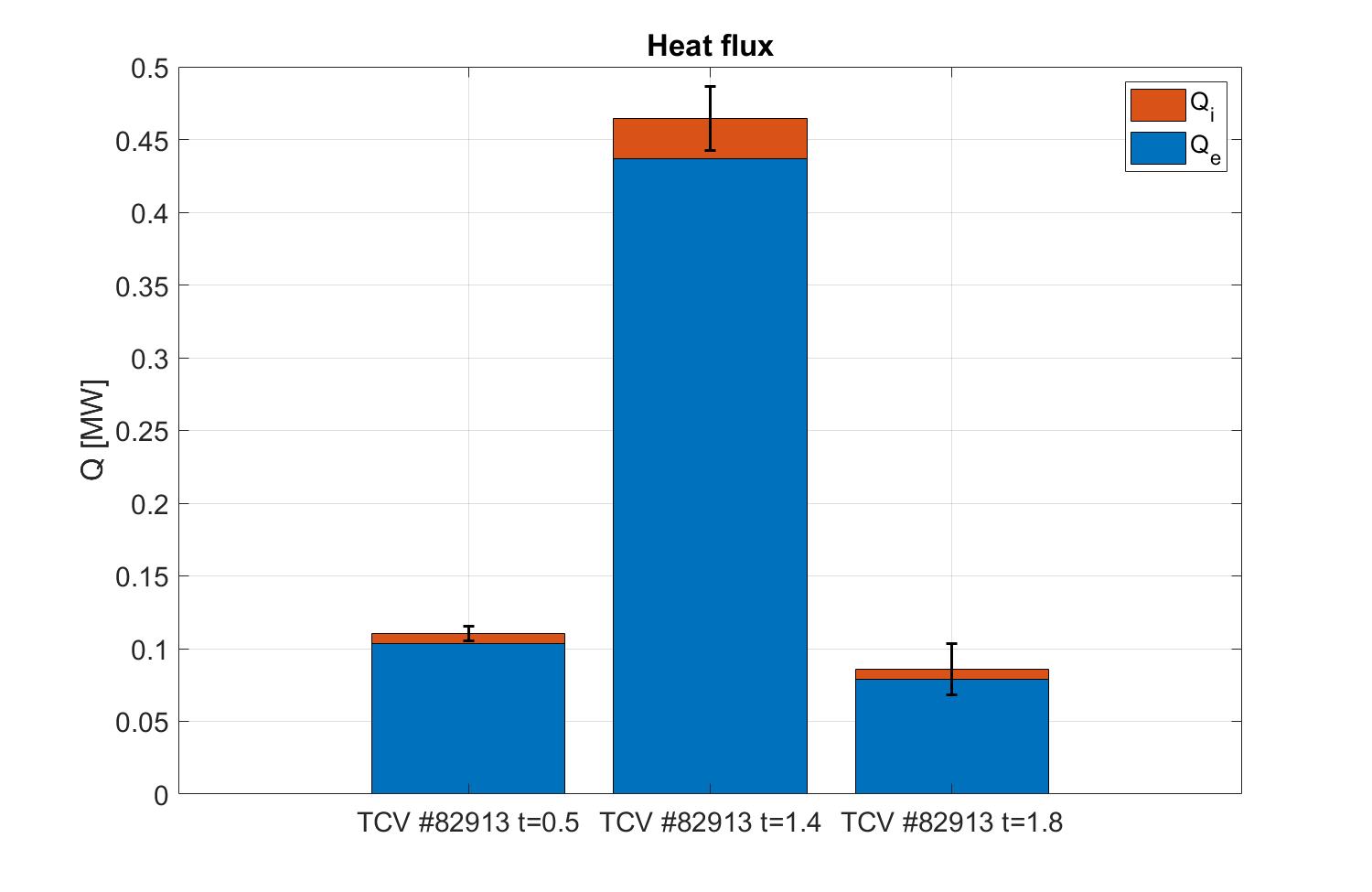}
    \caption{Electron electrostatic (blue) and ion electrostatic (red) component of the heat fluxes for the three considered scenarios. Error bars have been computed as standard deviations of the time traces of the heat fluxes.}
    \label{Q}
\end{figure}

\begin{figure}[h!]
    \centering
    \includegraphics[width=0.75\linewidth]{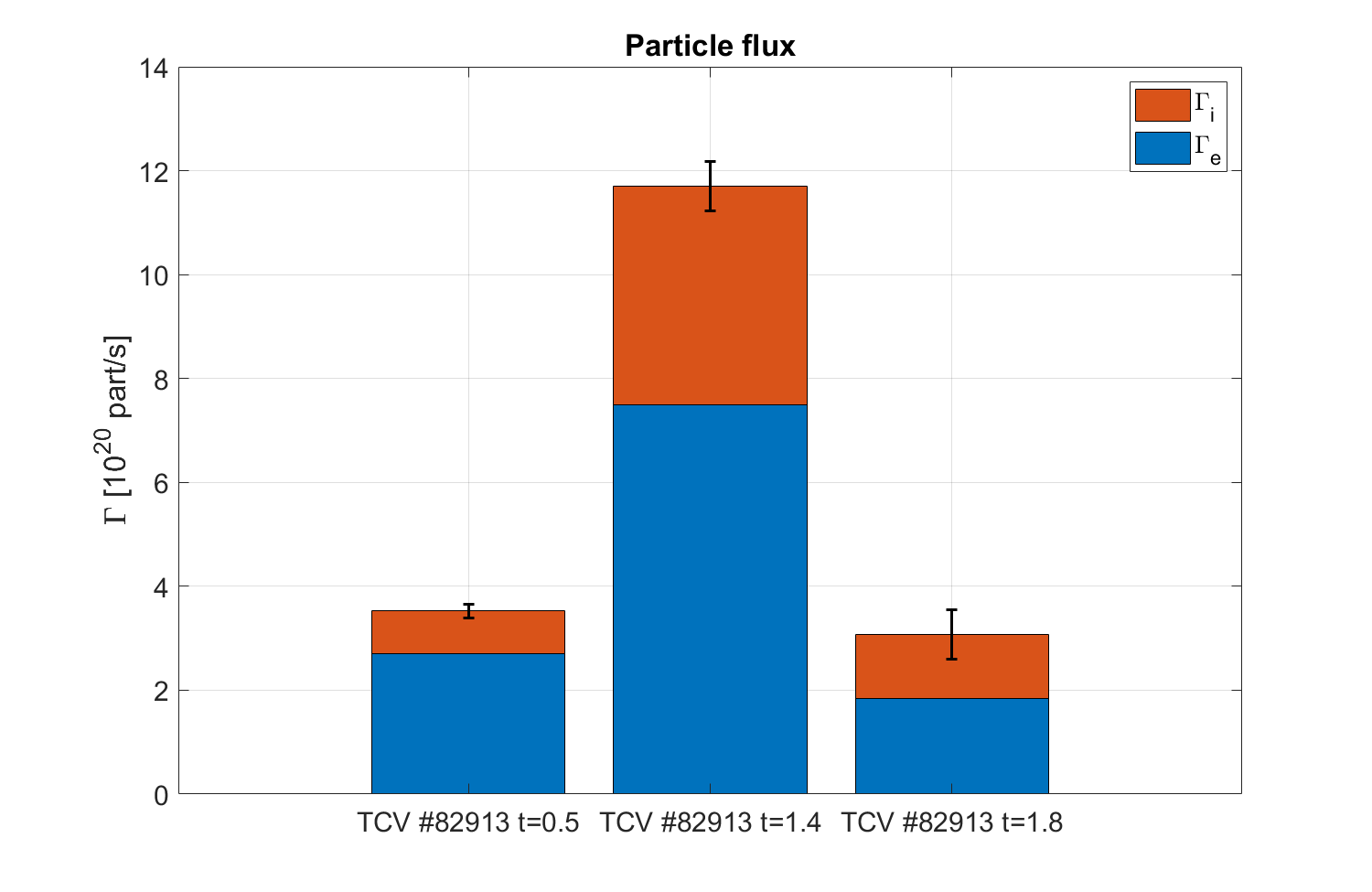}
    \caption{Electron electrostatic (blue) and ion electrostatic (red) component of the particle fluxes for the three considered scenarios. Error bars have been computed as standard deviations of the time traces of the particle fluxes.}
    \label{G}
\end{figure}

The enhanced instability of TEMs during the ECH phase is reflected in particle fluxes (Figure \ref{G}). At t=1.40 s, electron and ion particle fluxes are increased by a factor of 3 with respect to the two ohmic phases. This trend is coherent with the gas trace shown in Figure \ref{fig:ne core control 82913}. Indeed, when ECH is added, for the density level of $\mathrm{3.7 \cdot 10^{19} \ m^{-3}}$, a higher gas level has to be injected to maintain the central density constant at the target. This is due to stronger TEMs, which transport electrons and ions more effectively outside of the plasma. On the other end, when ECH is suddenly turned off at t=1.50 s, the central density increases rapidly. This change in the profile peaking is due to the stabilization of TEMs, which now transport fewer particles outside of the plasma due to the lack of the ECH-induced pumpout mechanism. Note that other GENE simulations have shown that with NBI heating the opposite to ECH is observed, with ITG and inward pinch dominating over TEM/outward pinch (see e.g. \cite{fontana_effects_2019}), which explains the peaked profile obatined with NBI, in Figure \ref{fig:ne profiles peaking 82893 82913}b.
\section{Control of the edge density in high density H-mode discharge}\label{section: ne edge control H-mode}

High-density H-mode operation is a leading strategy for fusion power plants such as ITER and DEMO, as it provides enhanced confinement and enables high fusion performance. Increasing plasma density improves fusion yield and promotes divertor detachment, ensuring sustainable heat and particle exhaust. However, high-density operation is constrained by the empirical density limit, which imposes an upper operational boundary. This limit is commonly expressed through the Greenwald fraction $\mathrm{f_{\mathrm{GW}}}$ \cite{Greenwald_1988}, defined as $\mathrm{n_{\mathrm{GW}}[10^{20} m^{-3}] = I_{p}[MA]/(\pi a^2[m^{2}])}$, $\mathrm{f_{\mathrm{GW}} = NEL/n_{\mathrm{GW}}}$, where $a$ is the plasma minor radius and $\mathrm{NEL}$ the line-averaged electron density, 
and is widely used as an indicator of proximity to density-limit disruptions and as a normalized performance metric.

More recently, refined descriptions of the density limit have been proposed, including a first-principles model \cite{Giacomin2022} and a data-driven multi-machine scaling \cite{Maris2025}, which provide improved or comparable accuracy to the Greenwald limit, particularly in reducing false positive predictions of density-limit disruptions. A comprehensive overview of the H-mode density limit is given in \cite{Manz2025_DensityRadiationLimits}. A strong dependency of the latter scalings and experimental observations on the plasma condition at the edge, such as edge density, collisionality, and power crossing the separatrix, suggests that accurate monitoring and control of the edge plasma quantities play a key role in a reliable and robust operation of fusion power plants in the high-density regime.
 
Different studies have been carried out to characterize the high-density H-mode regime in AUG and TCV tokamaks, especially identifying the density limits at different heating levels and plasma currents that initiate the H-to-L mode back-transition and plasma disruption. Real-time estimation of the density profile enables the monitoring of the plasma in these operational regimes and informs the actuator manager to provide an adequate response with the control system. In particular, in \cite{Bernert2015_HDL}, an extensive study on the different phases of H-mode entering conditions, sustainment, degradation, and loss of the high confinement regime has been carried out on AUG. The determination of a critical edge density, representing the highest density that can be achieved in H-mode before the degradation of confinement, has been conducted based on the line-averaged density measured by an interferometer that crosses the plasma edge. A scaling law can be determined from different shots designed to explore the density limit in various operational points, scanning the input gas fueling and heating power in the operational space.

The resulting empirical law \cite{Bernert2015_HDL} has been extracted:
\begin{equation}
    n_{e,\mathrm{crit.,edge}}=c_{ne} P_{\mathrm{heat}}^{c_{P}}I_{p}^{c_{I_{p}}} q_{95}^{c_{q_{95}}}
    \label{eq:critical edge density}
\end{equation}
\begin{equation}
    f_{\mathrm{crit.,edge}} = \dfrac{\bar{n}_{\mathrm{e,edge}}}{n_{e,\mathrm{crit.,edge}}}
    \label{eq:critical fraction edge density}
\end{equation}
   
Where $\mathrm{n_{e,\mathrm{crit.,edge}}[10^{20} m^{-2}]}$ is the critical density to be compared with the line-integrated density measured by an interferometer at the plasma edge $\mathrm{\bar{n}_{\mathrm{e,edge}}}$ (for AUG, the channel $H_{5}$), $\mathrm{P_{heat}[MW]}$ is the heating power injected in the plasma, Ip [MA] is the plasma current, $\mathrm{q_{95}}$ is the edge safety factor and $\mathrm{c_{ne}}$,  $\mathrm{c_{P}}$ and $\mathrm{c_{I_{p}}}$ are the fit coefficients which have been computed in the regression study in \cite{Bernert2015_HDL}, summarized in Table \ref{tab:coefficients ne crit edge}:

\begin{table}[H]
\centering
  \begin{tabular}{ccccc}
    \hline
    $c_{ne}$ & $c_{P}$ & $c_{I_{p}}$ & $c_{q_{95}}$ \\
    \hline
    $0.506$ & $0.396$ & $0.265$ & $-0.323$ \\
  \end{tabular}
  \caption{Coefficients employed to compute the critical edge electron density $n_{e,\mathrm{crit.,edge}}$ in Equation (\ref{eq:critical edge density}).}
  \label{tab:coefficients ne crit edge}
\end{table}

In \cite{Maraschek2018}, the tracking and control of the plasma confinement factor $\mathrm{H_{98,y2}}$ \cite{ITER1999_Ch2} and the normalized critical edge density $\mathrm{f_{\mathrm{crit.,edge}}}$ has been carried out, based on the critical density computed in Equation (\ref{eq:critical fraction edge density}) with the real-time data from the edge interferometer channel in ASDEX Upgrade. This control task is performed by modulating NBI input power and gas flux to provide operation at a given distance from a disruptivity boundary empirically determined in the space $\mathrm{H_{98,y2}-f_{\mathrm{crit.,edge}}}$.
A demonstration of the cross-machine compatibility of the disruption avoidance scheme has been carried out on TCV \cite{Vu_2021} and JET \cite{Sieglin2025_HDL_DisruptionAvoidance}, where similar control strategies tested on AUG have been applied to high-density-limit disruption avoidance shots, keeping the same disruptivity boundary for the $\mathrm{H_{98,y2}-f_{\mathrm{crit.,edge}}}$ plane and re-adjusting the measurements of the TCV or JET edge interferometer with the length of the FIR crossing the plasma LCFS. 

The operational experience gathered from these experimental results pinpoints new research paths and improvements to be undertaken in disruption avoidance control schemes.
Operating and performing accurate tracking and control of the edge density in such advanced scenarios is, in fact, a challenging task, due to different reasons:
\begin{itemize}
    \item High-density operation usually translates to a time-varying, non-negligible pick-up of the density in the SOL from the interferometry system \cite{Bernert2015_HDL}, especially if the channel crosses highly dense and radiative regions such as X-point radiators (XPR) \cite{Bernert2017} or MARFE \cite{Lipschultz1984}.
    \item The FIR signals are subject to fringe jumps in case of high-density plasmas, due to a reduced SNR. The signal loss is due to increased deflection of the interferometer laser in the high-density regions of the plasma \cite{Murari2006}. They can also result from density changes that exceed the diagnostic bandwidth and/or electromagnetic noise pickup \cite{Bosman2021}.
    \item The measurement of the edge density usually depends on a device-specific diagnostic system, hindering cross-machine comparison and the validity of scaling laws to be used for next step devices.
\end{itemize}
With the implementation of the new multi-rate electron density observer in the TCV PCS, a new set of experiments on proximity control and disruption avoidance has been carried out. The $n_e$ profile observer has been leveraged to inform the disruption avoidance controller deployed within SAMONE on SCD, with the required density information reconstructed in real time. In particular, synthetic FIR data produced from the RAPDENS forward diagnostics model and the reconstructed electron density profile are propagated to the controller.

In the framework of disruption avoidance and proximity control, RAPDENS $n_e$ observer can be leveraged to provide:
\begin{itemize}
    \item A reference to the controllers
    which filters fringe jumps in real time with the frequency of the TS system (see \ref{section:Appendix_scheme_observer}).
    \item FIR synthetic signals that account only for the density contained within the LCFS (see Section \ref{section:Detachment}). This aspect is particularly beneficial in the case of MARFE or XPR.
    \item A more refined reference signal for the edge density, computed by the multi-rate profile observer.
    \item A reference for the computation of $\mathrm{f_{\mathrm{crit.,edge}}}$, which is real-time capable and device-agnostic, where the edge electron density is mapped on magnetic flux surfaces and thus is independent of the relative position of the plasma and the device-specific diagnostics used to infer the edge density. 
\end{itemize}

To compute the critical edge density fraction $\mathrm{f_{\mathrm{crit.,edge}}}$ from RAPDENS reconstruction, two different approaches have been implemented: an average of the synthetic FIR signals $\#1$ and $\#2$ located on the LFS of TCV poloidal plane (see Figure \ref{fig:liuqe_TS_FIR}), using only the density inside the LCFS and the value of the $\mathrm{n_{e}}$ profile on the plasma pedestal, in this case computed as $\mathrm{n_{\mathrm{e,edge}}=n_{e}(\rho=0.70)}$. The first approach has been proven robust over multiple shots, since the reconstruction of the FIR signals is directly constrained by the EKF, due to the availability of FIR measurements in each RAPDENS particle state estimation, every 1 ms. The second approach fulfills the requirements in the last two points mentioned above. However, it has been proven reliable only if the convergence of the underlying predictive model is verified. In case of unsuccessful convergence of the predictive model, or in case of a nonphysical converged result, e.g. a negative value assumed by the plasma profile or a neutral particle inventory, the predicted value obtained in that specific EKF iteration is discarded and replaced by the last update of the corrected state by the EKF. The regularization of the profile reconstruction through the solution of the PDE in Equation (\ref{eq:density1d_eq}) is then lost or indirectly applied through the TS-corrected FIR signals used in the EKF update step (for more details, see \ref{section:Appendix_scheme_observer}).

In the remaining part of this Section, an example of disruption avoidance experiments using RAPDENS FIR averaged edge density and a high-performance H-mode shot with direct control of $\mathrm{f_{\mathrm{crit.,edge}}}$ computed from $\mathrm{n_{\mathrm{e,edge}}=n_{e}(\rho=0.70)}$ are presented.

\subsection{Disruption avoidance control, $\mathrm{n_{\mathrm{e,edge}}}$ computed from RAPDENS synthetic FIR}

Figure \ref{fig:DA 80525, plot H98 vs f_crit} represents the trajectory of the TCV shot \#80525 in the $\mathrm{H_{98,y2}-f_{\mathrm{crit.,edge}}}$ plane. The black solid line determines the separation between the high-disruptivity region (enclosed in the bottom-right region of the plot) and the safe operational space in these normalized units. Notably, the disruptivity boundary is identical to the one adopted for AUG \cite{Maraschek2018} and JET \cite{Sieglin2025_HDL_DisruptionAvoidance} disruption avoidance experiments, making this work valuable for cross-machine comparison. In the presented shot, an indirect control of the trajectory in the $\mathrm{H_{98,y2}-f_{\mathrm{crit.,edge}}}$ plane has been carried out by actively controlling plasma $\mathrm{\beta_{\mathrm{tor}}}$ and $\mathrm{f_{\mathrm{crit.,edge}}}$ through, respectively, control of NBI-1 power and gas valve fueling, as reported in Figure \ref{fig:DA 80525,control task} . The evolution of the H-mode phase of the shot, in the time interval $\mathrm{t\in[1.00 \ s, 1.50 \ s]}$, starts from a low density fraction of $\mathrm{f_{\mathrm{crit.,edge}}\approx0.60}$ (blue trace of Figure \ref{fig:DA 80525, plot H98 vs f_crit}), to a recovery phase (in green) at around t=1.15 s where a minimum distance of $\mathrm{d_{H98,y2-f_{\mathrm{crit.,edge}}}\approx0.15}$ between the shot trajectory and the disruptivity boundary is reached, finally ending in an increased confinement of $\mathrm{H_{98,y2}\approx1.25}$ and critical density friction of $\mathrm{f_{\mathrm{crit.,edge}}\approx1.1}$ (in dark red).

\begin{figure}[h!]
\centering
  \includegraphics[width=0.65\textwidth]{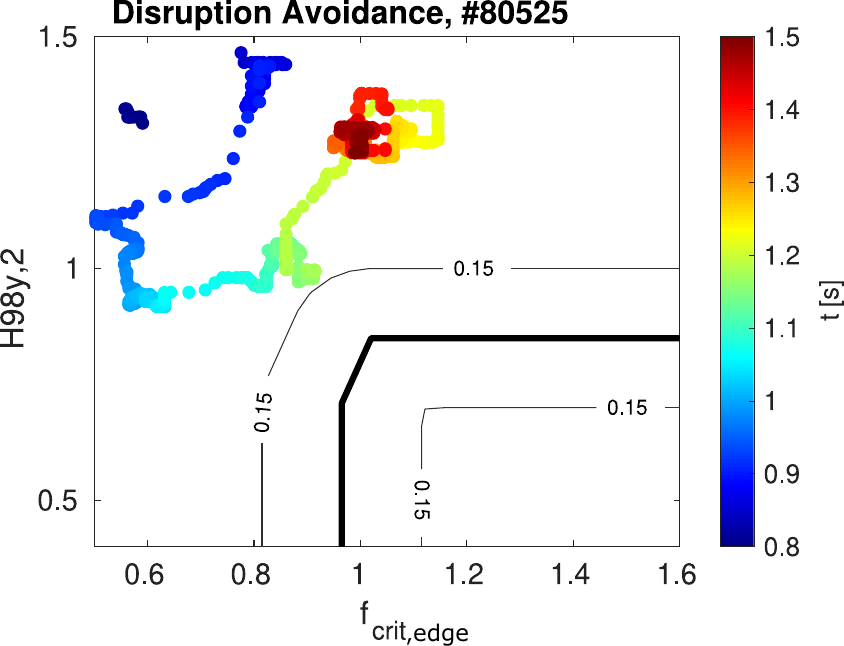}
  \centering
  \caption{Trajectory of the shot in the plane $H_{98,y2}-f_{\mathrm{crit.,edge}}$ for the disruption avoidance control experiment on TCV, shot \#80525. The solid line in black represents the boundary between the operational region and the high disruptivity region, enclosed on the bottom-right of the plot. The trajectory of the shot (represented in time from blue to red) represents a recovery scenario, where a minimum in the distance $\mathrm{d_{H98,y2-f_{\mathrm{crit.,edge}}}}$ is reached at t=1.15 s, after which an increasing value of both $\mathrm{H_{98,y2}}$ and $\mathrm{f_{\mathrm{crit.,edge}}}$ is reached by increasing NBI power and gas valve fueling.}
  \centering
  \label{fig:DA 80525, plot H98 vs f_crit}
\end{figure}

\begin{figure}[h!]
\centering
  \includegraphics[width=0.95\textwidth]{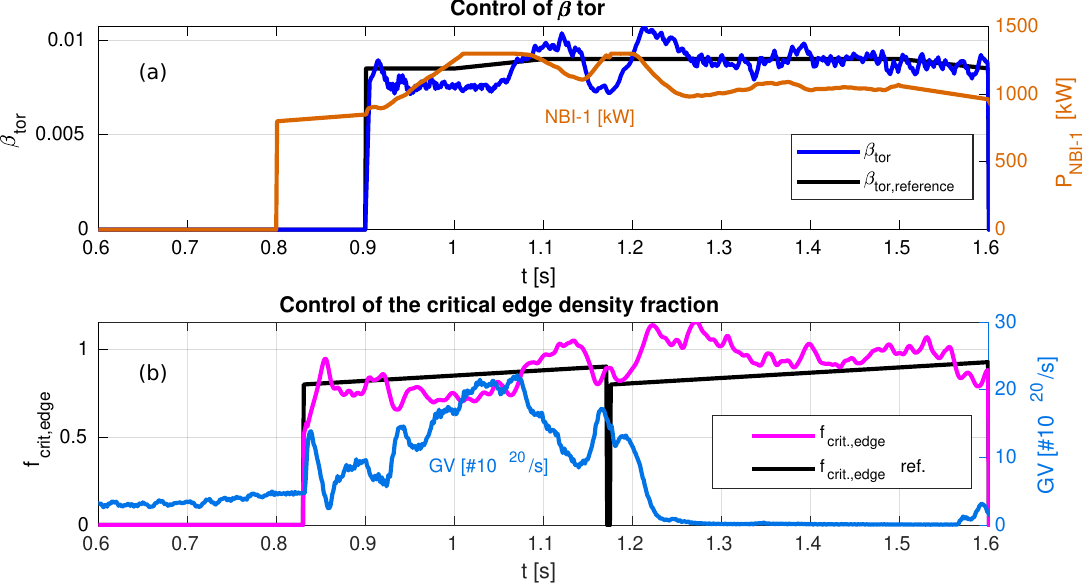}
  \centering
  \caption{Simultaneous control of $\beta_{\mathrm{tor}}$ and critical edge density fraction in shot \#80525. In Figure \ref{fig:DA 80525,control task}a, the reference $\mathrm{\beta_{tor,ref.}}$(in black) and the controlled $\mathrm{\beta_{\mathrm{tor}}}$ (in blue) are reported, together with the feedback signal for the injected power from the NBI-1 (in orange). In Figure \ref{fig:DA 80525,control task}b, the density control performance is presented, with the reference critical edge density fraction $\mathrm{f_{\mathrm{crit.,edge}} \ ref.}$(in black), the controlled $\mathrm{f_{\mathrm{crit.,edge}}}$ (in magenta) and the gas valve flow (in cyan) are shown.}
  \centering
  \label{fig:DA 80525,control task}
\end{figure}

 Figure \ref{fig:80525_ne_edge_DA} shows some quantities computed in real time by the multi-rate observer. In the first subplot, a comparison of the RAPDENS reconstructed FIR computed within the LCFS, named $FIR_{\mathrm{LCFS}}$, for the edge channels \#1 and \#2 and the rescaled value of $n_{e}(\rho=0.70)$ is reported. A remarkable agreement between the two quantities can be verified, highlighting the validity of using the FIR average within the LCFS of the external channels as a proxy for the edge density computed in Equation (\ref{eq:critical fraction edge density}). The validity stands provided that no shape modification or radial position movement occurs during the shot, as the relative position between the interferometer channels and plasma shape would change. This provides a first improvement with respect to previous experiments, including TCV \cite{Vu_2021}.
 
 The Figures \ref{fig:80525_ne_edge_DA}b and Figures \ref{fig:80525_ne_edge_DA}c show the RAPDENS reconstruction of FIR channels \#8 and \#9 compared with the digitally filtered channels, which have been affected by fringe jumps. The comparison of the RAPDENS reconstructed FIR signal and the synthetic signal computed using the fitted TS profile on the FIR \#8 line of sight in the last subplot showcases the agreement of the reconstruction.
 
 These results highlight the capabilities of the observer to successfully reconstruct the synthetic data signal from FIR channels not directly employed in the EKF reconstruction (since only channels from \#1 to \#6 have been actively incorporated in the EKF procedure). 

\begin{figure}[h!]
\centering
  \includegraphics[width=0.75\textwidth]{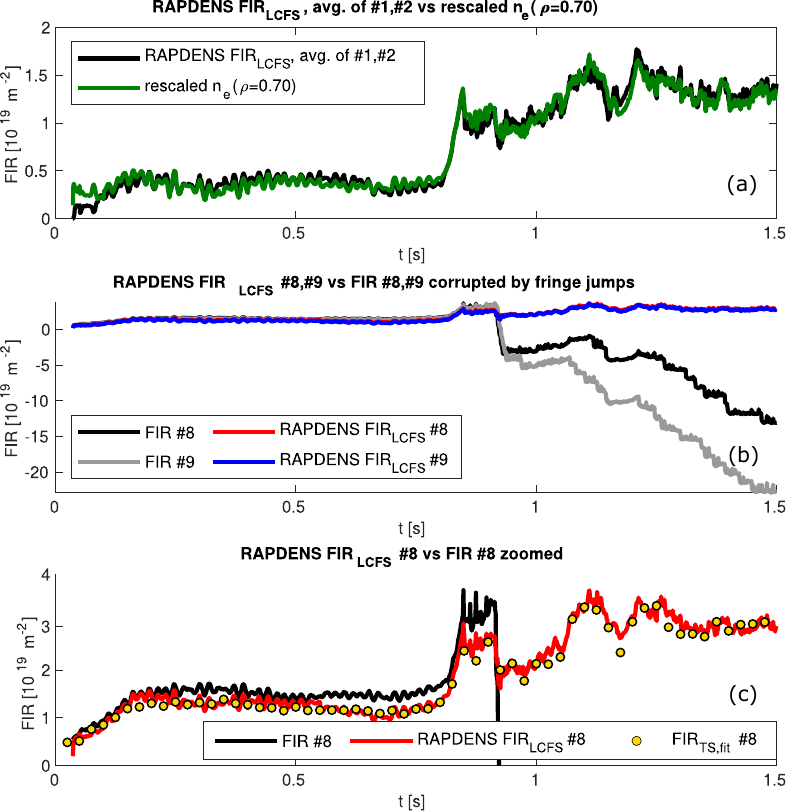}
  \centering
  \caption{Synthetic FIR channels reconstruction with RAPDENS for shot \#80525. The average of the FIR channels \#1 and \#2 has been used as a proxy to compute the edge density for the critical density fraction $f_{\mathrm{crit.,edge}}$. Figure \ref{fig:80525_ne_edge_DA}a shows the comparison between RAPDENS $FIR_{\mathrm{LCFS}}$ (in black) and the rescaled density reconstructed with the observer at $n_{e}(\rho=0.70)$ (in green). Figures \ref{fig:80525_ne_edge_DA}b and \ref{fig:80525_ne_edge_DA}c demonstrate the observer's ability to reconstruct faulty FIR signals in real time, which are corrupted by fringe jumps. FIR signals \#8 and \#9 are reported in black and gray, together with the reconstructed synthetic signals from the observer, in red and blue. On Figure \ref{fig:80525_ne_edge_DA}c the reconstructed signal for channel \#8, in red, is compared with offline fitted TS profiles remapped on the FIR \#8 line of sight and spatially integrated, in yellow, and the corrupted FIR signal, in black.}
  \centering
  \label{fig:80525_ne_edge_DA}
\end{figure}

\subsection{High-performance, high-density H-mode shot, $n_{\mathrm{e,edge}}$ computed from RAPDENS $n_{e}(\rho=0.70)$} \label{subsection:H-mode_edge_ne_rho_0.7}

In this Section, a high-density, high-performance H-mode shot with feedback control of the critical density fraction computed with RAPDENS using $n_{e}(\rho=0.70)$ is reported. A high $\beta_{N}$ in the flat-top of around $\beta_{N}\approx2.15$, has been achieved, in conjunction with a Greenwald fraction of $f_{\mathrm{GW}}\approx0.80$. Simultaneous control of the critical density fraction $f_{\mathrm{crit.,edge}}$ and toroidal beta $\beta_{\mathrm{tor}}$ has been carried out with two independent PID controllers, modulating the gas inlet flux and the power of NBI-1. Figure \ref{fig:82876_high_performance} shows the time traces that describe the evolution of the main plasma quantities during the experiment. In Figure \ref{fig:82876_high_performance}a, the plasma current $I_p$, in black, evolves from the flattop value of $-140 \ kA$ up to successful ramp-down \cite{wang_learning_2025}. The negative value of the current is related to TCV coordinates conventions, such that the $I_p$ current is aligned with the NBI-1 injection direction. 
Figure \ref{fig:82876_high_performance}b shows the injected NBI-1 power, in orange, the reference for $\beta_{\mathrm{tor}}$, in black, the controlled trace, reconstructed with RT-LIUQE, in blue, and the ELMs frequency, in green. In this high-density regime, the ELMs are quite small in amplitude and characterized by a higher frequency compared with typical type-I ELMs \cite{Labit2019_SmallELM_TCV_AUG}. In Figure \ref{fig:82876_high_performance}c, the reference $f_{\mathrm{crit.,edge}}$, in black, the $f_{\mathrm{crit.,edge}}$ computed with the density reconstructed with RAPDENS using $n_{e}(\rho=0.70)$, in magenta, and the gas flux GV injected to control it, in light blue, is shown.
The two control tasks are executed simultaneously in SAMONE, and a good control of the two quantities can be observed within the time window $\mathrm{t=[0.900 \ s, 1.30 \ s]}$. In the subsequent section of the shot, feedforward commands to NBI-1 power and gas valve fueling input are introduced with a pre-programmed shutdown task inserted through SAMONE.

In Figure \ref{fig:82876_high_performance}d, the reconstructed density with the observer, in black, is compared with offline TS data, in red, and the averaged value of FIR channels \#1 and \#2, in light green. A good temporal agreement between the reconstructed signal and the TS point fitted on the radial normalized coordinate $\rho=0.70$ can be appreciated. The density trace is not only coherent with the TS points at $40 \ Hz$, since only two TS lasers out of three were available, but it follows the evolution of the density in the L-to-H transition and ramp-down, with a reconstruction frequency of 1 kHz. A numerical inaccuracy in the reconstruction can be noticed at $\mathrm{t = [0.950 \ s,0.980 \ s]}$, due to a failure in convergence of the predictive model in such a time window. The effect is visible, leading to an underestimation of the density at $\rho=0.70$. As the new TS point in time is available, at t = 0.975 s, the convergence of the predictive model is achieved, with the adoption of the newly estimated $\mathrm{\nu_{\mathrm{TS}}/D}$ (as described in Section \ref{section:estimation_nu_TS}). A remarkable difference in the H-to-L back-transition and ramp-down between the RAPDENS reconstructed time trace $\mathrm{n_{e}(\rho=0.70)}$ and the average of FIR \#1,\#2 can be observed. This mismatch can be attributed to the motion of the plasma in the poloidal plane on the HFS, which occurs in the time $\mathrm{t = [1.30 \ s,1.40 \ s]}$, before the transition diverted-to-limited takes place at t=1.40 s. This example shows how the density reconstruction can be decoupled from the radial position control with the RAPDENS observer. The mapping of the density profile on flux surfaces enables an improved estimation of the edge density compared with the sole use of the FIR channels, due to the time-varying change of the intersection between the LFS edge FIR chords \#1 and \#2 and plasma radial position. We thereby demonstrate that using the edge density allows to use the proximity to density limit phase space throughout the whole discharge, including ramp-down and significant plasma shape changes.


\begin{figure}[h!]
\centering
  \includegraphics[width=0.75\textwidth]{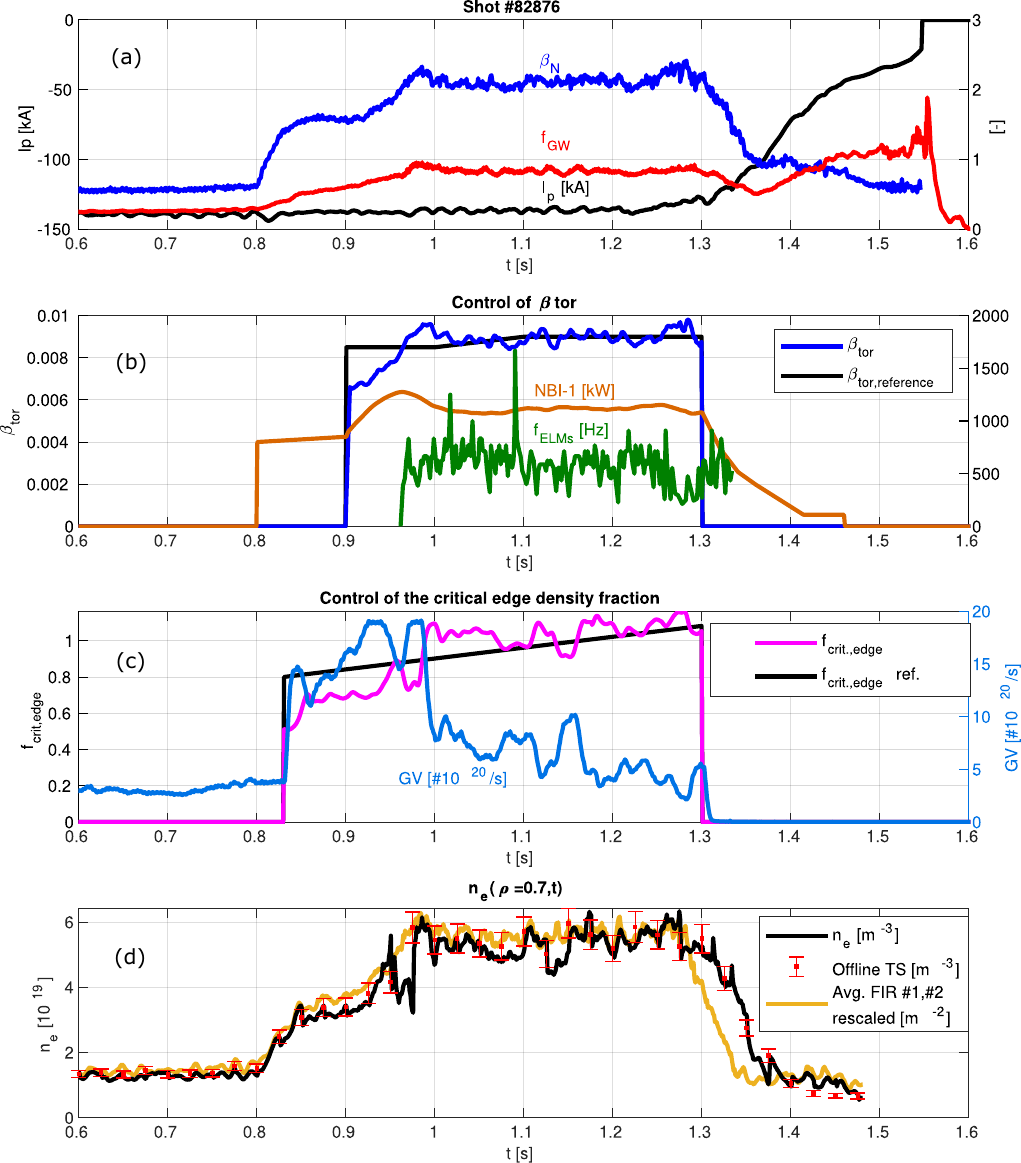}
  \centering
  \caption{High density, high performance  H-mode plasma, TCV shot \#82876. High normalized beta of $\beta_{N}\approx2.15$ and Greenwald fraction of $f_{\mathrm{GW}}\approx0.80$ are achieved (Figure \ref{fig:82876_high_performance}a). Simultaneous control of $\beta_{\mathrm{tor}}$ control with NBI-1 (Figure \ref{fig:82876_high_performance}b) and edge density fraction through gas puffing (Figure \ref{fig:82876_high_performance}c), leveraging RAPDENS reconstruction of $n_{e}(\rho=0.70)$, is successfully demonstrated. ELMs frequency, in green, and NBI-1 power, in orange, are shown in the same scale in Figure \ref{fig:82876_high_performance}b. The reconstructed time trace of the density at $n_{e}(\rho=0.70)$, in black, well aligns with the offline TS points, in red, throughout the different sections of the shot in Figure \ref{fig:82876_high_performance}d. A large disagreement between the rescaled average of the FIR channels \#1,\#2 and the RAPDENS reconstructed time trace at $n_{e}(\rho=0.70)$ can be noted during the plasma ramp-down, due to the motion of the plasma in the poloidal plane towards the HFS.}
  \centering
  \label{fig:82876_high_performance}
\end{figure}

\section{Conclusion \& Outlook}\label{section: conclusion}

In this work, an improved multi-rate RAPDENS observer, employed for the reconstruction of the electron density profile on TCV, is presented and tested with a series of experiments relevant for future power plants, namely with high SOL density, high $\beta_N$ near the density limit, robust local density control near ECH cut-off, and profile control with heating mix. The observer incorporates high-spatial resolution Thomson scattering data with a high temporal resolution far-infrared interferometer in an Extended Kalman Filter approach. It includes, as well, a real-time adaptation of the model parameter used for the profile prediction. The latter allows reliable density profiles despite missing diagnostics or fringe jumps. The observer, coupled with a gas flux PI controller embedded in the supervisory actuator manager, SAMONE, is tested over various experimental scenarios. The novel density control scheme enabled the control of the line-averaged electron density within the LCFS for the control of the upstream density conditions in support of detachment studies, as presented in Section \ref{section:Detachment}. The scheme ensures reproducible upstream density conditions in a divertor-agnostic fashion, overcoming the limitations of the traditional density control scheme, especially linked to the non-negligible pickup of the SOL density in the divertor region, in high-density and/or high impurity seeding conditions.

Local density control of the electron density profile has been demonstrated in various experimental conditions, encompassing ECH and NBH plasmas in L-mode plasmas and high-density, high-performance H-mode plasmas. In Section \ref{section: ne control below cutoff}, the employment of the RAPDENS-based control of the central electron density in ECH/NBH shots has been presented, providing a reference to the gas flux controller, which enables the precise control of the density in the deposition location of the ECH power below cutoff conditions, coherent with offline Thomson scattering data. The time-varying relationship between the central FIR channel and the central density value highlights how the observer can provide a more accurate reference to the density controller, removing the dependency of the density profile shape and peaking factor from the signal to be controlled. A qualitative examination of the different profiles in the heating mixes applied during the experiments has been conducted, showing how the profile peaking factor is influenced by the action of the ECH and NBI power at constant central density. A quantitative analysis has been additionally carried out with the GENE code to confirm the cause of the modifications of the particle transport in the ECH shot \#82913. Consistent with what is reported in similar studies, such as \cite{Weisen2001}, the anomalous transport in this regime is TEM-dominated. The resulting pumpout effect of ECH on the plasma is reflected in a flattened profile.

Finally, simultaneous control of the critical edge density fraction and $\mathrm{\beta_{\mathrm{tor}}}$ in high-density, high-performance H-mode plasmas is showcased. The controllers act independently on the gas valve and NBI power to maintain the performance of the plasma stable, or to investigate a disruption avoidance scenario, extending proximity to density limit control at high performance to a normalized density tokamak and measurement-type agnostic. The observer provides a reference to the density controller, which is device-agnostic (non-dependent on the specific diagnostics used to measure the edge density for a given device) and consistent with Thomson scattering offline data. An example of reconstruction of fringe-jump corrupted signals with a subset of FIR channels is also presented, highlighting the observer's capability to reconstruct the faulty channels with the forward diagnostics model.

Different research paths can stem from this work. So far, the profile reconstruction capability of the observer has been proven reliable for a subset of radial normalized toroidal coordinates, i.e. $\rho\in[0,0.80]$. To further extend the reconstruction capabilities of RAPDENS EKF, the adoption of the inhomogeneous Dirichlet boundary conditions for the real-time reconstruction is envisaged, as presented in \cite{Kropackova2025} and shown in the offline example reported in \ref{section:Appendix_inhomog_BC}.

On the density control side, the RAPDENS predictive model can be leveraged to incorporate additional physics, such as a dependency of the pinch velocity-to-diffusivity ratio $\nu/D$ on the injected ECH power, to reproduce the flattened density profile which results from a central power deposition due to ECH, or vice versa, the peaked density profile resulting from NBI heating due to the ionization of the injected neutrals in the plasma core, while both heat source affect (opposite) the particle transport properties.

The improved model can be used as a basis to design an MPC, such as the one designed in \cite{BosmanITER}, for MIMO profile density control, or, generally, to systematically design the density controller by means of a predictive model simulator that embeds experimental information, such as the estimated $\nu/D$ with the EKF as shown in Equation (\ref{eq:nu_eq}).

The estimation of other unknown parameters, or the explicit adoption of disturbances in the EKF for the reconstruction of the density state or for offsets between FIR and TS diagnostics, as shown in \cite{VanMulders2025}, is considered to further improve the predictive model capabilities and to generalize the procedure in the correction of diagnostics data within the multi-rate formulation of the EKF problem. 

A step forward, starting from the work done in Section \ref{section:Detachment}, is to couple the novel density controller with an exhaust observer/controller scheme. The estimation of the profile can be used in conjunction with the exhaust scheme to design a MIMO controller as done in \cite{Koenders2023_MIMOController_TCV}. In such a way, precise control of upstream conditions and divertor detachment can be achieved, for improved simultaneous control of core-edge-exhaust, in view of integrated control for future fusion power plants.

The interfacing of RAPDENS with other state reconstruction algorithms, such as RAPTOR or TORBEAM \cite{POLI200190}, in the TCV PCS is envisioned for improved real-time state estimation of electron and ion temperatures, safety factor profile, deposition location, and absorption of EC power and current drive. The estimated kinetic profiles can inform a pressure profile controller for real-time control of the pedestal pressure in H-mode plasmas, or for the feedback control of internal transport barriers in advanced scenarios \cite{Sauter2005_ITBProbe}.

Improved real-time Kinetic Equilibrium Reconstruction \cite{Carpanese2020}, coupling the estimates of the pressure and current density profiles to the RT-LIUQE equilibrium magnetic solver, is finally envisaged for a self-consistent solution of the Grad-Shafranov equation. The solution of the inverse, free-boundary equilibrium problem is aided by means of the kinetic observers that incorporate internal profile measurements in the solution. In this way, accurate reconstruction of the q-profile and the location of internal flux surfaces is enabled, for monitoring and control of the safety factor profile in hybrid and advanced scenarios \cite{piron_extension_2019} and suppression of MHD modes.

\section{Acknowledgements}
This work has been carried out within the framework of the EUROfusion Consortium, partially funded by the European Union via the Euratom Research and Training Programme (Grant Agreement No 101052200 — EUROfusion).
The Swiss contribution to this work has been funded in part by the Swiss State Secretariat for Education, Research and Innovation (SERI).
Views and opinions expressed are however those of the author(s) only and do not necessarily reflect those of the European Union, the European Commission or SERI. 
Neither the European Union nor the European Commission nor SERI can be held responsible for them.

Data employed for the generation of the presented results are available upon reasonable request to the main corresponding author.
\appendix
\section{Details on the structure of the multi-rate observer}\label{section:Appendix_scheme_observer}

The schematic of the electron density observer, collocated in the control scheme  presented in Section \ref{section:Control scheme SCD}, Figure \ref{fig:SCD_scheme}, is here reported. The processing of the data, with and without the real-time availability of TS data, is differentiated with green and red line connectors.

\begin{figure}[h!]
\centering
  \includegraphics[width=\textwidth]{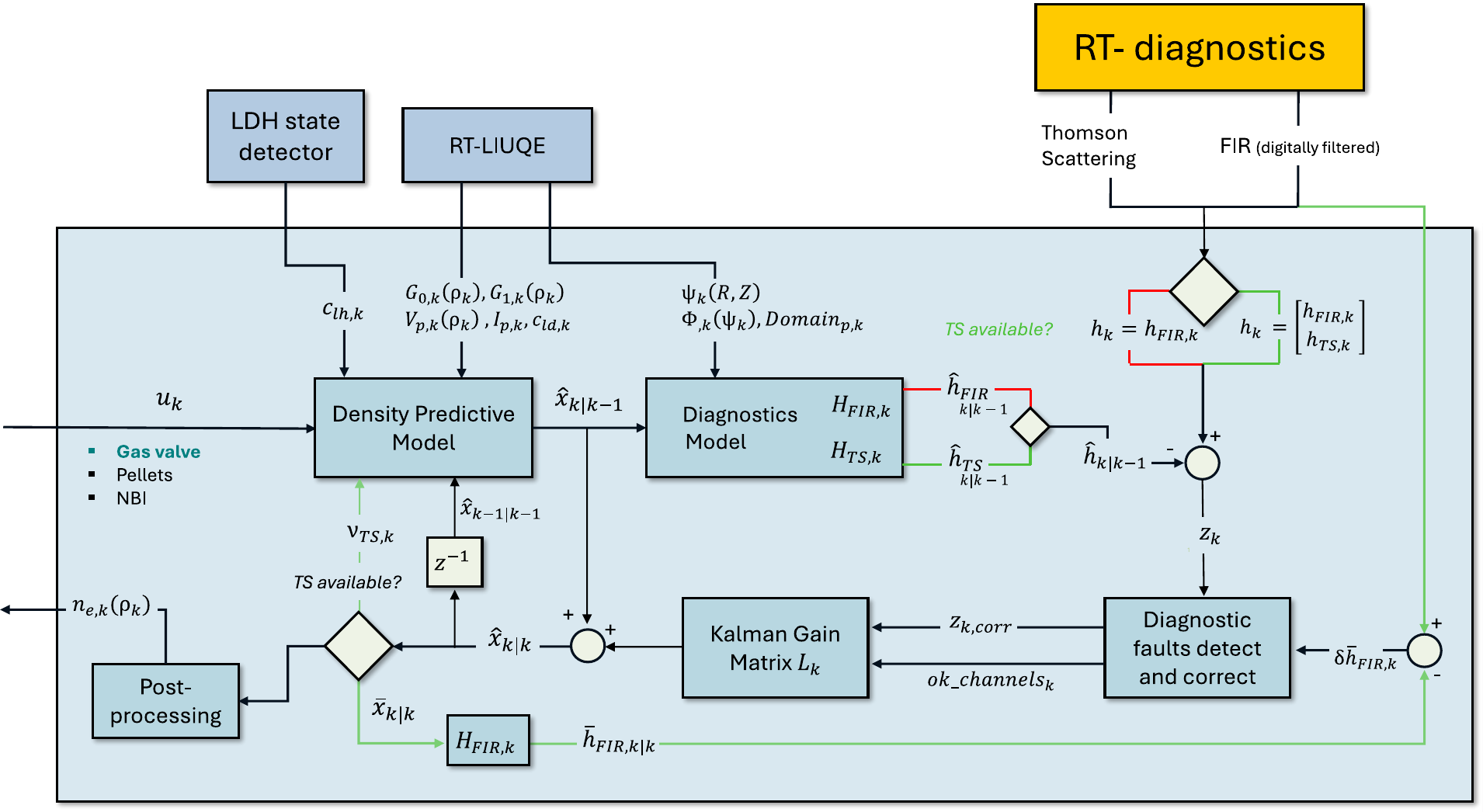}
  \centering
  \caption{Schematic of the new RAPDENS multi-rate density observer.}
  \centering
  \label{fig:rapdens_scheme}
\end{figure}

Starting from the top-left corner of Figure \ref{fig:rapdens_scheme}, the \textit{Density Predictive Model} block takes as input the signals $u_{k}$, which contain the amount of fueling particles injected in the vacuum chamber from the gas valve, the boolean $c_{lh,k}$, provided from the LDH state detector, which indicates whether the plasma is in L- or H-mode, the magnetic equilibrium information from RT-LIUQE, i.e. the plasma volume $V_{p,k}$, current $I_{p,k}$ and geometry profiles $G_{0,k}$ and $G_{1,k}$, the updated state estimated on the previous time step $\hat{\textbf{x}}_{k-1|k-1}$ and the electron pinch velocity $\nu_{\mathrm{TS},k}$ computed with Equation (\ref{eq:nu_eq}), when TS data are available.

The output of the \textit{Density Predictive Model}, $\hat{\textbf{x}}_{k|k-1}$, represents the prediction step described in Equation (\ref{eq:predictive_eq}).
RT-LIUQE provides additional equilibrium information for the assembly of the forward \textit{Diagnostics Model} defined in Equation (\ref{eq:synth_meas_eq}), such as the plasma poloidal flux map $\psi_k=\psi_k(R,Z)$, the toroidal flux $\Phi_k=\Phi_k(\psi_k)$ defined as a function of the poloidal flux $\psi_k$, and a logical mask that determines the position of the plasma up to the LCFS in the poloidal plane, denominated $Domain_{p,k}=Domain_{p,k}(R,Z)$. This latter signal has been employed to incorporate in real time the subset of TS points enclosed up to the LCFS.

The estimation of the synthetic measurement of the FIR signal $\hat{\textbf{h}}_{\mathrm{FIR}_{k|k-1}}$ is performed and propagated at each timestep, every 1 ms; meanwhile, the estimated synthetic measurement of the TS data, $\hat{\textbf{h}}_{\mathrm{TS}_{k|k-1}}$, is stacked in the synthetic diagnostic estimate $\hat{\textbf{h}}_{k|k-1}$ only when new TS data are available. 

The innovation residual $\textbf{z}_{k}$ is computed with the available measurement information contained in $\textbf{h}_{k}$ and processed in the \textit{Diagnostic faults detect and correct} block. In this block, a correction, applied as an additive offset $\delta\bar{\textbf{h}}$, corrects the FIR channels based on the estimate of the updated $\bar{\textbf{h}}_{k|k}$ provided by the EKF procedure when TS data are available. With this step, offsets related to the presence of density outside the LCFS, or channels affected by fringe jumps, are successfully corrected and propagated to the update step of the EKF algorithm.

In this final section, the updated state $\hat{\textbf{x}}_{k|k}$ is computed by means of Equation (\ref{eq:correction_step}), multiplying the corrected innovation residual $\textbf{z}_{k,\mathrm{corr}}$ to the \textit{Kalman Gain Matrix} $L_{k}$, while taking into account only the available diagnostic channels, stored in $ok\_channels_k$. The state $\hat{\textbf{x}}_{k|k}$ is then passed to a time shift block $z^{-1}$ for the next predictive model iteration and a \textit{Post-processing} block for the computation of the density profile $n_{e,k}=n_{e,k}(\rho)$ and derived quantities, such as the updated synthetic measurements for FIR $\hat{\textbf{h}}_{\mathrm{FIR}_{k|k}}$ and TS $\hat{\textbf{h}}_{\mathrm{TS}_{k|k}}$ diagnostics.

In case of availability of TS data in real-time, the updated state, labeled $\bar{\textbf{x}}_{k|k}$, is used to compute the electron pinch velocity $\nu_{\mathrm{TS},k}$, as reported in Equation (\ref{eq:nu_eq}), and the estimate of the FIR synthetic data $\bar{\textbf{h}}_{\mathrm{FIR}_{k|k}}$ to be used in the \textit{Diagnostic fault detect and correct} block. The values of $\nu_{\mathrm{TS},k}$ and $\bar{\textbf{h}}_{k|k}$ are kept constant with a zero-order hold and updated as new TS data are available in the SCD framework.
\section{Offline application of the inhomogeneous Dirichlet boundary condition for the reconstruction of the density profile}\label{section:Appendix_inhomog_BC}
In Section \ref{section:ne_profile_heating_schemes}, Figure \ref{fig:ne profiles TS 82893 82913}, it has been highlighted how the RAPDENS version adopted in the experiments for TCV shots \#82913 and \#82893 suffers from an underestimation in the reconstruction of the electron density profile in the radial region located at $\rho\geq0.80$. A key point that affects the reconstruction of the density profile in this radial location is the adoption of homogeneous Dirichlet boundary conditions, located at the end of the numerical grid, where the SOL is located.

One possible way to overcome such a limitation is to provide inhomogeneous boundary conditions to the numerical solver and restrict the profile estimation inside the LCFS, at $\rho=1.0$. In \cite{Kropackova2025}, an upgraded version of the RAPDENS code has been tested and validated with offline AUG data, introducing a non-zero Dirichlet boundary condition at the plasma separatrix. The separatrix density has been estimated using an empirical formula \cite{Kallenbach2018}, which scales with the divertor neutral pressure and can be applied in real-time using ionization gauge data measurements. The results shown in \cite{Kropackova2025} provide a remarkable improvement in the reconstruction of the profile at the edge, especially if the online estimation of the pinch-to-diffusivity transport coefficient $\nu/D$ is carried out with the formula reported in Equation (\ref{eq:nu_eq}). In the considered case, the estimation of the unknown transport coefficient has been conducted with interferometer signals in the EKF reconstruction.
As a remark, the ionization particle source term and recombination and SOL sink terms have been deactivated in this upgraded model, since the core particle dynamics and SOL region have been numerically decoupled. The interfacing of the code with a control-oriented model for the SOL has been successfully carried out in \cite{Kropackova_thesis}, using as a basis for the implementation the multi-inventory, reduced particle-balance model presented in \cite{Muraca2023}.

The upgraded RAPDENS code has been adopted for TCV, recently tested with the reconstruction of the electron density profile, while simultaneously estimating the separatrix density and transport coefficients by means of the spatially resolved RT-TS data. Tuning of the upgraded RAPDENS predictive model for offline and real-time applications is underway, for local control of upstream density conditions and plasma pedestal, around $\rho=0.90$. 

In this Appendix, a comparison between the real-time data provided in shot \#82913 with the RAPDENS homogeneous Dirichlet boundary conditions for the edge and an offline reconstruction of the density profile using RAPDENS inhomogeneous boundary conditions is shown. The offline simulation employs the same input data used for the real-time reconstruction. In Figures \ref{fig:timetraces_82913_homog_vs_inhomog} and \ref{fig:timeslices_inhomog_vs_homog_82913}, the performance of the two versions of the code is reported against offline TS data, for radial coordinates $\rho\in[0.70,0.80,0.90,1.0]$ and times $t\in[0.300\ s,0.800 \ s, 1.20 \ s, 1.80 \ s]$, reported in Figure \ref{fig:timetraces_82913_homog_vs_inhomog} with dashed, magenta vertical lines. 

\begin{figure}[h!]
\centering
  \includegraphics[width=0.75\textwidth]{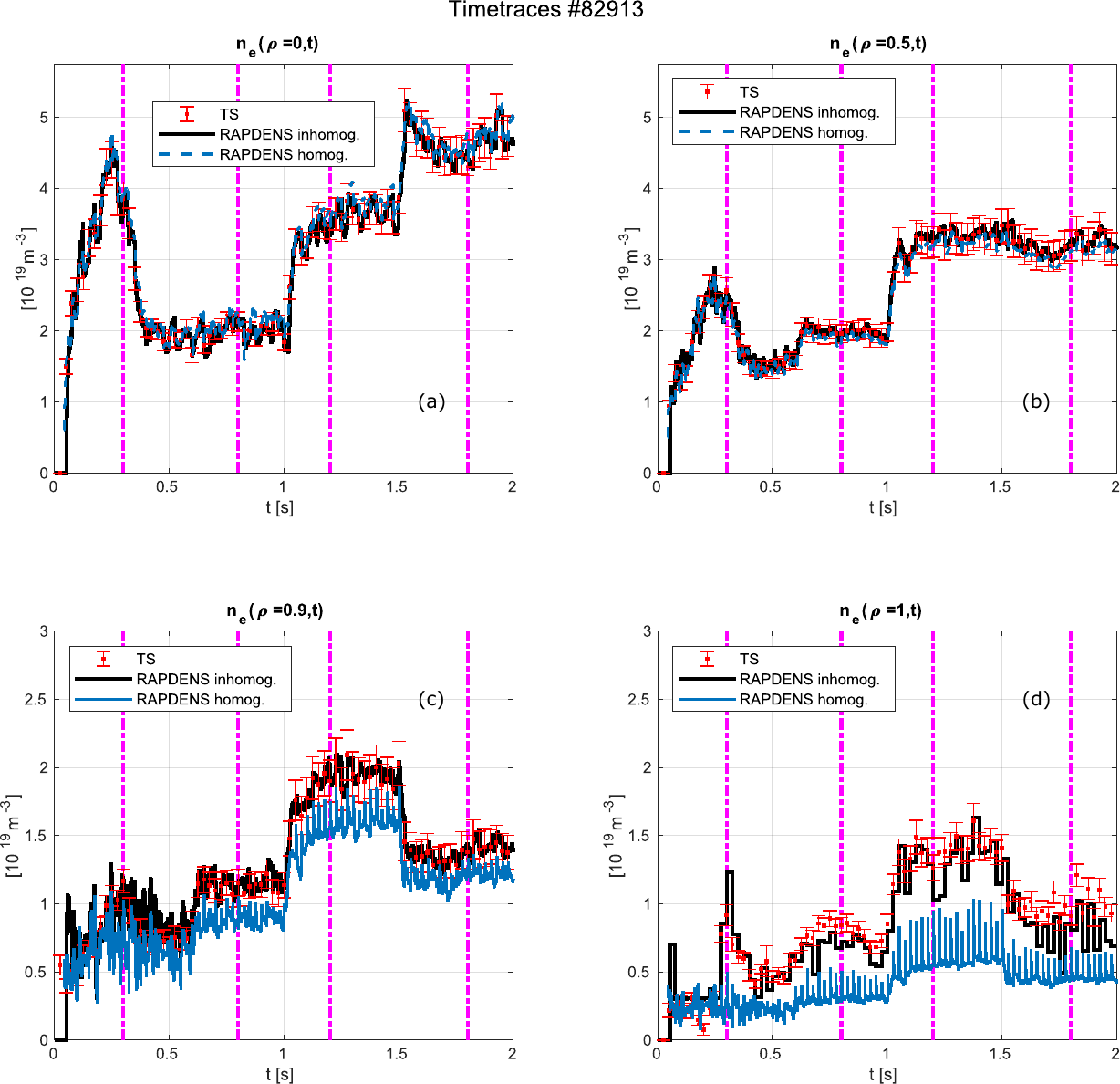}
  \centering
  \caption{Comparison of RAPDENS Dirichlet homogeneous boundary conditions (in cyan) and RAPDENS Dirichlet inhomogeneous (in black) against TS data (in red), shot \#82913. Different timetraces at the following radial locations $\rho\in[0.70,0.80,0.90,1.0]$ are shown.}
  \centering
  \label{fig:timetraces_82913_homog_vs_inhomog}
\end{figure}

\begin{figure}[h!]
\centering
  \includegraphics[width=0.75\textwidth]{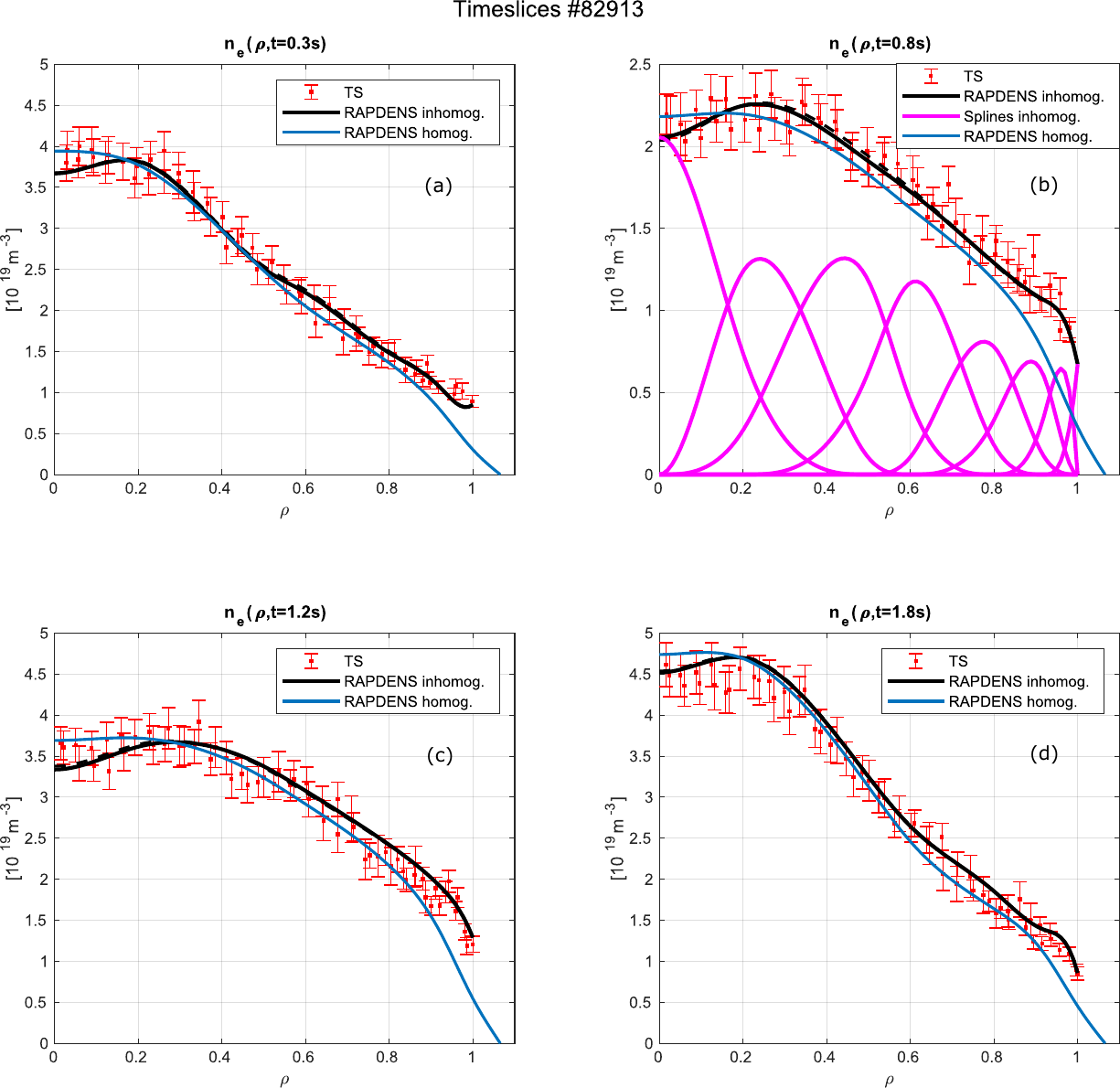}
  \centering
  \caption{Comparison of RAPDENS Dirichlet homogeneous boundary conditions (in cyan) and RAPDENS Dirichlet inhomogeneous (in black) against TS data (in red), shot \#82913. Different timeslices are selected in correspondence to the magenta vertical lines reported in Figure \ref{fig:timetraces_82913_homog_vs_inhomog}, in limited, diverted, ECH, and ohmic phases of the shot.}
  \centering
  \label{fig:timeslices_inhomog_vs_homog_82913}
\end{figure}

A remarkable improvement in the reconstruction of the density near the edge can be appreciated. Notably, since no source or sink terms are present in the Dirichlet inhomogeneous version of the code, the approximation done in Equation (\ref{eq:nu_eq}) for the computation of $\nu_{\mathrm{TS}}/D$, where the source term $\hat{S}$ is omitted, is coherent with the numerical solution of the particle transport equation. 

The timetraces in cyan for $n_{e}(\rho=0.90)$ and $n_{e}(\rho=1.0)$ highlight the effect of the TS data on the local estimation of the profile, appearing as sudden spikes in the timetraces. The information coming from the TS is not sufficiently retained either in the prediction step of the EKF (with the adapted $\nu_{\mathrm{TS}}/D$) nor in the correction step (with the adjusted value of the FIR using the offsets $\delta \bar{h}_{\mathrm{FIR},k}$, as explained in \ref{section:Appendix_scheme_observer}). Conversely, in the inhomogeneous Dirichlet boundary condition case, a higher accuracy in the reconstruction is evident. The mismatch in the estimation of the separatrix density $n_{e}(\rho=1.0)$ in the time interval $t=[0.100 \ s,0.200\ s]$ is related to a fault in the estimation of the separatrix density, due to a missed assignment of RT-TS data at the separatrix. Ongoing work is currently carried out to improve the estimation in such cases, with fallback options in case of faulty measurements and integration of neighboring TS points close to the separatrix to increase the information available in this plasma region.  
\clearpage
\section{Covariance matrices adopted in the RAPDENS multi-rate Extended Kalman Filter}\label{section:Appendix_covariance_matrices}
The state $Q_{k}$ and measurement $R_{k}$ covariance matrices introduced in Section \ref{section:summary_RAPDENS_EKF_equations} are here reported for completeness. It is highlighted that these tuning coefficients for the EKF have been kept constant for the different shots reported in this article, and how the single pair of $(Q_{k},R_{k})$ enables a good performance of the observer in various experimental scenarios and density time dynamics. In the remainder of this Appendix, the structure and tuning choices of the covariance matrices are briefly summarized. The state covariance matrix $Q_{k}$ is structured as a symmetric Toeplitz matrix, and is set constant over time. This structure has been adopted as it is from the work done in \cite{Blanken2018}. The first 8 diagonal elements, shown in Figure \ref{fig:covariance_matrix_Q} and, more in detail, in Figure \ref{fig:Qk_plots_diag_offdiag}, represent the values of the covariance to the spline coefficients used to numerically discretize the density profile; meanwhile, the last two are the covariance values referring to the wall and vacuum inventories.

\begin{figure}[h!]
\centering
  \includegraphics[width=0.45\textwidth]{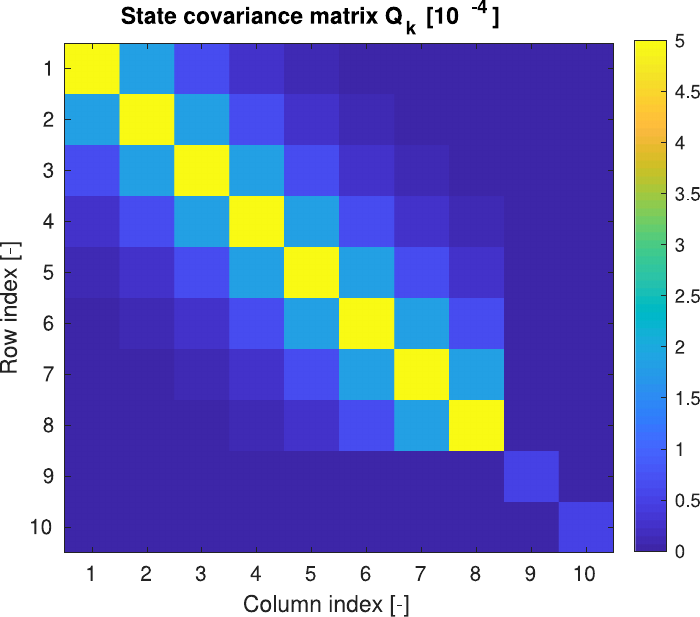}
  \centering
  \caption{State covariance matrix $Q_{k}$ in normalized units, scaled $[10^{-4}]$.}
  \centering
  \label{fig:covariance_matrix_Q}
\end{figure}

\begin{figure}[h!]
\centering
  \includegraphics[width=0.75\textwidth]{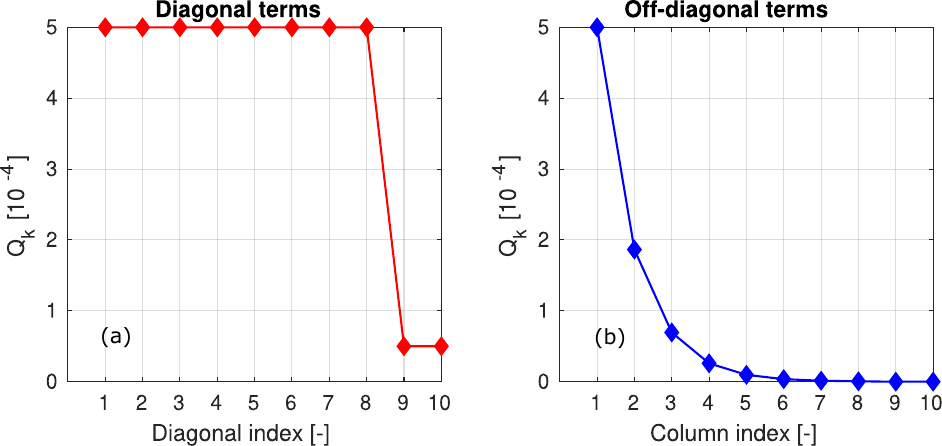}
  \centering
  \caption{Detail of the matrix $Q_{k}$, showing the diagonal (in red) and off-diagonal terms (in blue). The exponential decay of the off-diagonal elements is highlighted in the right subplot. Normalized units, scaled $[10^{-4}]$.}
  \centering
  \label{fig:Qk_plots_diag_offdiag}
\end{figure}

The off-diagonal covariance elements, reported in Figure \ref{fig:Qk_plots_diag_offdiag}b, represent the spatial correlation between a given spline state and the neighboring ones. Increasing this spatial correlation, by setting a different lower decay coefficient, results in smoother profiles, but decreases the time response of the EKF to local variations of the profile shape. Notably, no covariance has been assigned to the wall and vacuum inventory states, thus treating them as statistically independent states. 
It is also possible to change the relative weight between the different spline coefficients, for example, by introducing lower variance to the innermost splines that represent the core of the profile, and keep a higher covariance for the estimation of the pedestal/edge region through the diagnostic information. An in-depth investigation of these tuning choices is envisaged, especially concerning the estimation of the separatrix density with the EKF, as mentioned in \ref{section:Appendix_inhomog_BC}. 

The measurement covariance $R_{k}= diag(R_{k,FIR},R_{k,\mathrm{TS}})$ represents the expected noise associated with each diagnostic channel in the EKF. In this specific case, the variance values of each channel are set and remain constant across all diagnostic channels. The chosen values for the associated standard deviations of FIR and TS diagnostics are $\sigma_{\mathrm{FIR}}=\sigma_{\mathrm{TS}}=8.0\cdot10^{-3}$, resulting in a measurement covariance matrix of $R_{k}(i,i)=6.4\cdot10^{-5}$, computed with Equation (\ref{eq:measurement covariance_eq}). The choice of setting $\sigma_{\mathrm{FIR}}=\sigma_{\mathrm{TS}}$ is the result of manually tuning the measurements' covariances, thereby weighting the effect of each diagnostic noise in the reconstruction. Small changes in this value between the two diagnostics would result in a shift of the reconstruction towards the one having the lowest covariance. 

The values of the matrix $R_{k}$ have been set one order of magnitude lower than the values present on the diagonal term of $Q_k$, indicating that the observer heavily relies on the measurement data for the state estimation. In particular, the chosen covariance matrices result in a near-instantaneous fitting of the TS data as the information is available in real time. The profile is then used to update the pinch velocity transport coefficient $\nu_{\mathrm{TS}}$ and the FIR-to-TS offset $\delta \bar{h}_{\mathrm{FIR}}$, as explained in \ref{section:Appendix_scheme_observer}.

Shifting the weight towards the model, by decreasing the covariance of $Q_{k}$ while keeping the $R_{k}$ entries fixed, is considered in the case of good model predictive capabilities, for example can be implemented when repeating similar discharges. Furthermore, in the case of the availability of an estimate of error bars in real time for FIR or TS, the information can be translated into a time-varying expression of the covariance measurements matrix $R_{k}$, excluding or reducing the weight of chords or data points with large error bars or flagged as corrupted.  
\clearpage

\bibliographystyle{model1-num-names}
\bibliography{NF25.bib}

\end{document}